\documentclass[titlepage, 12pt]{article}

\usepackage{authblk}
\usepackage[margin=1in]{geometry}
\usepackage[linesnumbered,ruled,vlined]{algorithm2e}
\usepackage{amsbsy, amsfonts,
amsmath, amssymb, amsthm}

\usepackage{multirow}
\usepackage{xcolor}
\usepackage{adjustbox}
\usepackage{tikz}
\usetikzlibrary{arrows.meta}

\usepackage{dsfont}

\usepackage{bm}

\usepackage{booktabs, threeparttable}
\usepackage{setspace}

\usepackage{natbib}

\usepackage{color}

\definecolor{ForestGreen}{RGB}{34,139,34}

\usepackage{graphicx}
\graphicspath{{./}, {../figures/}}
\usepackage{physics}
\usepackage{subcaption}

\usepackage[colorlinks=true, citecolor=blue]{hyperref}
\usepackage{mathrsfs}

\allowdisplaybreaks 

\newcommand{\bfv}{\bm{v}}

\newcommand{\bfq}{\bm{q}}
\newcommand{\dout}{d^{\, (1)}}
\newcommand{\din}{d^{\, (2)}}
\newcommand{\pout}{p^{\, (1)}}
\newcommand{\pin}{p^{\, (2)}}
\newcommand{\qout}{q^{\, (1)}}
\newcommand{\qind}{q^{\, (2)}}
\newcommand{\tq}{\tilde{q}}
\newcommand{\bftq}{\bm{\tilde{q}}}
\newcommand{\indicator}{\mathbb{I}}
\newcommand{\bftheta}{\bm{\theta}}
\newcommand{\deltain}{\delta_{\, {\rm in}}}
\newcommand{\deltaout}{\delta_{\, {\rm out}}}
\newcommand{\bfM}{\bm{M}}
\newcommand{\bfk}{\bm{k}}
\newcommand{\bfl}{\bm{l}}
\newcommand{\bfR}{\bm{R}}
\newcommand{\bfS}{\bm{S}}
\newcommand{\bfI}{\bm{I}}
\newcommand{\bfK}{\bm{K}}
\newcommand{\bfL}{\bm{L}}
\newcommand{\ER}{\mathrm{ER}}

\let\proglang=\textsf \let\code=\texttt

\begin{document}

\title{An Efficient Algorithm for Generating Directed Networks with 
Predetermined Assortativity Measures}

\author[1]{Tiandong Wang}
\affil[1]{Department of Statistics, Texas A\&M University, College Station, TX 77843}
\author[2]{Jun Yan}
\affil[2]{Department of Statistics, University of Connecticut,
  Storrs, CT 06269}
\author[2]{Yelie Yuan}
\author[3]{Panpan Zhang}
\affil[3]{Department of Biostatistics, Epidemiology and
  Informatics, University of Pennsylvania, Philadelphia, PA 19104}

\maketitle

\doublespacing

\begin{abstract}
	Assortativity coefficients are important metrics to analyze
	both directed and undirected networks. In general, it is not
	guaranteed that the fitted model will always agree with the 
	assortativity 
	coefficients in the given network, and the structure of directed 
	networks is more
	complicated than the undirected ones. Therefore, we provide a 
	remedy
	by proposing a degree-preserving rewiring 
	algorithm, called DiDPR, for generating directed networks with 
	given directed assortativity coefficients. We 
	construct the joint edge 
	distribution of the target network by accounting for the four 
	directed assortativity coefficients simultaneously, provided 
	that 
	they are attainable, and obtain the desired network by solving a 
	convex optimization problem.
	Our algorithm also helps check the attainability of the given 
	assortativity coefficients. We assess the performance of the 
	proposed algorithm by simulation studies with focus on two 
	different network models, namely Erd\"{o}s--R\'{e}nyi and 
	preferential attachment random networks. We then apply the 
	algorithm to a Facebook wall post network as a real data 
	example. The codes for implementing our algorithm are publicly 
	available in \proglang{R} package 
	\code{wdnet}~\citep{Rpkg:wdnet}.
	
\end{abstract}

\section{Introduction}
\label{sec:intro}

Assortativity is a metric measuring the tendency that nodes in a 
network are connected to vertices with similar
node-specific characteristics, such as node degrees and 
strengths~\citep{Newman2002assortative, Yuan2021assortativity}.
Originally, \citet{Newman2002assortative} proposed an assortativity 
measure based on node degrees for unweighted, undirected networks. 
Analogous to Pearson's correlation coefficient, assortativity 
coefficient ranges from $-1$ to $1$, with a positive (negative) 
value indicating that high degree nodes are likely to be connected 
with high (low) degree nodes. A network with positive (negative) 
assortativity coefficient is called assortative (disassortative) 
mixing. 

For directed networks, \citet{Newman2003mixing} defined an 
assortativity measure based on 
out-degrees of source nodes and in-degrees of target nodes.
In general, there are four types of assortativity 
measures for directed networks~\citep{Foster2010edge, 
	Piraveenan2012assortative}, namely 
out-in, out-out, in-in and in-out assortativity coefficients. For 
instance, a large positive out-in assortativity coefficient suggests 
that 
source nodes with large out-degrees tend to link to target nodes 
with large in-degrees. Assortativity has 
been extended to weighted,
undirected networks \citep{Leung2007weighted} and weighted, directed
networks \citep{Yuan2021assortativity}. For other recent 
developments, see \citet{Chang2007assortativity, Holme2007exploring, 
	Litvak2013uncovering, Noldus2015assortativity}.

Generating random networks with given assortativity
is of great theoretical and practical interest. 
On one hand, once a hypothesized model has been fitted to a given 
dataset,
checking whether the fitted network has achieved the assortativity 
levels 
from the original dataset is an important way to assess the goodness 
of fit.
On the other hand, if discrepancies on the assortativity 
coefficients have been 
observed, one can propose necessary improvements, e.g. edge 
rewiring, 
for the hypothesized model to
better capture the underlying network dynamics. 
For undirected networks, such attempts have been made in 
\citet{Newman2003mixing}, where a 
degree-preserving rewiring algorithm has been proposed. The crux
of the algorithm is to 
construct a target network with the given 
assortativity coefficient, then rewire the initial network towards 
the target. \citet{Newman2003mixing} achieved this goal
by characterizing the distribution for the 
number of edges connecting two nodes with certain degrees,
but this method is unfortunately inapplicable to directed networks, 
especially when there are four assortativity coefficients to control 
simultaneously.

Recently, \citet{Bertotti2019configuration} developed a rewiring 
method for obtaining the maximal and minimal assortativity 
coefficients 
in undirected networks. However, limited work has been done for 
directed networks. 
\citet{Kashyap2017mechanism} introduced a rewiring algorithm 
focusing on only one of the four assortativity coefficients 
in directed networks, but overlooked the other three. 
\citet{Uribe2021finding} proposed a three-swap method to investigate 
the profile of rank-based assortativity measures in directed 
networks. To the best of our knowledge, there is no research 
accounting for four types of assortativity coefficients and their 
attainability simultaneously in directed networks. 
Hence, one of the primary goals of the present paper is 
to fill this gap.

Here we propose a feasible, 
efficient rewiring algorithm for directed networks towards 
the four given assortativity levels simultaneously, 
provided that they are attainable. Our algorithm is both a 
complement 
and a generalization of the two-swap degree-preserving 
algorithm in \citet{Newman2003mixing}, 
hereafter referred to as Newman's algorithm. The incompatible 
component in
Newman's algorithm, i.e., the construction of joint edge 
distributions, is handled by formulating and solving a 
convex optimization problem, from which we further generalize and 
extend the 
algorithm to account for the four assortativity 
coefficients simultaneously. 
After a certain number of rewiring attempts, all assortativity 
coefficients in the resulting networks will attain their target 
values. In addition, since the four types of assortativity 
measures are dependent on each other, after fixing 
the value of one of the four assortativity measures, not all values 
in $[-1, 1]$ can be reached for the other three. Our algorithm then 
provides 
legitimate bounds for the coefficients, which has not been 
considered in Newman's algorithm. 
The implementation of the proposed algorithm is in an open-source
\proglang{R} package \code{wdnet} \citep{Rpkg:wdnet}.

The rest of this paper is organized as follows. The proposed 
algorithm is presented with a discussion on legitimate bounds of the
predetermined assortativity levels in Section~\ref{sec:convexprog}.
An extensive simulation study is reported
in~\ref{sec:sim}, where two widely used random network 
models, i.e. the Erd\"{o}s-R\'{e}nyi and the preferential attachment 
models, are considered. 
in Section~\ref{sec:appl}, we also apply the proposed algorithm to 
a real dataset obtained from the Facebook wall post network.
Important discussions, concluding remarks and extensions are then 
provided in Section~\ref{sec:dis}.

\section{Rewiring towards Given Assortativity Coefficients}
\label{sec:convexprog}

We start with a review on directed assortativity coefficients in
Section~\ref{sec:dassort}, after which we present our rewiring 
algorithm in Section~\ref{sec:rewiring}. We also look into the 
bounds of directed assortativity coefficients in 
Section~\ref{sec:range}.

\subsection{Directed Assortativity}
\label{sec:dassort}

Let $G = G(V, E)$ be a network with node set $V$ and 
edge set $E$. For any $v_1, v_2 \in V$ that are connected, we use 
$(v_1, v_2) \in E$ to represent a directed edge from source node 
$v_1$ to target node $v_2$. As our goal is to study directed 
networks, we simplify the notations by calling out- and in-degree 
type~1 and~2, respectively. We use $\dout_v$ and $\din_v$ to denote 
the out- and in-degrees of node $v \in V$, respectively. Let 
$\indicator(\cdot)$ be standard indicator function. Given a network 
$G(V, E)$, define the empirical out-degree and in-degree 
distributions respectively as
\begin{equation}
	\label{eq:def_pinpout}
	\pout_k 
	:=\frac{1}{\lvert V\rvert} \sum_{v \in V} \indicator(\dout_{v} = 
	k),
	\qquad
	\pin_l := \frac{1}{\lvert V\rvert}
	\sum_{v \in V} \indicator(\din_{v} = l), \qquad k,l\ge 0,
\end{equation}
where $\lvert V\rvert$ denotes the cardinality of the node set $V$. 
Let $e^{(a,b)}_{kl}$ 
be the proportion of edges from a source node of type $a$ degree $k$ 
to a target 
node of type $b$ degree $l$ for $a, b \in \{1, 2\}$. We use $q$ and 
$\tq$ to distinguish the marginal distributions for source and 
target nodes, respectively. For instance, $q_k^{(a)} := \sum_{l} 
e_{kl}^{(a,b)}$ refers to the probability that an edge emanates from 
a source node of type $a$ degree $k$, whereas $\tq_l^{(b)} := 
\sum_{k} e_{kl}^{(a,b)}$ is the probability that an edge 
points to a target node of type $b$ degree~$l$. The four types of 
directed assortativity coefficients~\citep{Yuan2021assortativity} 
are given by 
\begin{equation}
	\label{eq:4dassort}
	r(a, b) = 
	\frac{\sum_{k,l} kl \left(e^{(a,b)}_{kl} - q_k^{(a)} 
		\tq_l^{(b)}\right)}{\sigma_q^{(a)} \sigma_{\tq}^{(b)}},
	\qquad a,b \in \{1,2\},
\end{equation} 
where 
\[
\sigma_q^{(a)} = \sqrt{\sum_k k^2 q_k^{(a)} - \left(\sum_k k 
	q_k^{(a)}\right)^2} \; \mbox{and} \; \sigma_{\tq}^{(b)} = 
\sqrt{\sum_l l^2 \tq_l^{(b)} - \left(\sum_l l \tq_l^{(a)}\right)^2}
\] 
are standard deviations of $q_k^{(a)}$ and $\tq_l^{(b)}$, 
respectively.

Before presenting our algorithm, we need to define a few more 
notations. Let $\nu_{kl}$ be the proportion of nodes with 
out-degree $k$ and in-degree $l$. By 
\eqref{eq:def_pinpout}, we have
\[
\pout_k = \sum_{l} \nu_{kl} \qquad \mbox{and} \qquad \pin_l = 
\sum_{k} \nu_{kl}.
\]
Define also $\eta_{ijkl}$ as the proportion of directed 
edges linking a source node with out-degree~$i$ and in-degree 
$j$ to a target node with out-degree $k$ and in-degree $l$. Then the 
following relations hold, which are the building blocks for the 
development 
of our algorithm:
\begin{align}
	\label{eq:restr1}
	\sum_{j, l} \eta_{ijkl} = e^{\left(1, 1\right)}_{ik}, 
	&\qquad
	\sum_{j, k} \eta_{ijkl} = e^{\left(1, 2\right)}_{il}, 
	\qquad
	\sum_{k, l} \eta_{ijkl} = \frac{i \, \nu_{ij}}{\sum_{i, j} i 
		\,\nu_{ij}},
	\\ \label{eq:restr2}
	\sum_{i, l} \eta_{ijkl} = e^{\left(2, 1\right)}_{jk}, 
	&\qquad 
	\sum_{i, k} \eta_{ijkl} = e^{\left(2, 2\right)}_{jl}, 
	\qquad
	\sum_{i, j} \eta_{ijkl} = 
	\frac{l \, \nu_{kl}}{\sum_{k, l} l\, \nu_{kl}}.
\end{align}
Additionally, we write $q$
and $\tq$ as functions of $\eta_{ijkl}$ and $\nu_{kl}$:
\begin{align*}
	\sum_{j, k, l} \eta_{ijkl} = \frac{\sum_j i \,
		\nu_{ij}}{\sum_{i, j} i \, \nu_{ij}} = \qout_i, 
	&\qquad 
	\sum_{i, k, l} \eta_{ijkl} = \frac{\sum_i i \,
		\nu_{ij}}{\sum_{i, j} i \, \nu_{ij}} = \qind_j,
	\\
	\sum_{i, j, l} \eta_{ijkl} = \frac{\sum_l l \,
		\nu_{kl}}{\sum_{k, l} l \, \nu_{kl}} = \tq_k^{(1)}, 
	&\qquad 
	\sum_{i, j, k} \eta_{ijkl} = \frac{\sum_k l \,
		\nu_{kl}}{\sum_{k, l} l \, \nu_{kl}} = \tq_l^{(2)}.
\end{align*}
Hence, by
Equation~\eqref{eq:4dassort}, all assortativity coefficients, 
$r(a,b)$, 
$a,b\in \{1,2\}$, are functions of $\nu_{kl}$ and $\eta_{ijkl}$.
This is a crucial observation which helps develop our
degree-preserving rewiring algorithm in Section~\ref{sec:DiDPR}.

\subsection{Rewiring Algorithm for Directed Networks}
\label{sec:rewiring}

We start with a succinct review of Newman's 
algorithm~\citep{Newman2003mixing} for undirected networks, followed 
by proposing our algorithm for directed networks. The proposed 
algorithm is a non-trivial extension of Newman's algorithm.

\subsubsection{Newman's Algorithm for Undirected Networks}
Let $G_0(V, E)$ be the initial undirected network, and $r^* \in [0, 
1]$ be 
the target assortativity coefficient. For undirected networks, if 
there is an 
edge connecting $v_1, v_2 \in V$, we denote it by $\{v_1, v_2\} \in 
E$. In addition, we use $e_{kl}$ to represent the 
proportion of 
edges connecting two nodes respectively with degree $k$ and $l$. The 
pseudo codes of Newman's algorithm are given in 
Algorithm~\ref{alg:newman}, for which a sufficiently large $T \in 
\mathbb{N}$ is required to ensure convergence of the algorithm.

\begin{algorithm}[tbp]
	\caption{Pseudo codes for Newman's Algorithm.}
	\label{alg:newman}
	\KwIn{Initial network $G_0(V, E)$, number of rewiring steps $T$, 
		target assortativity coefficient $r^*$.}
	\KwOut{$G(V, E)$.}
	Compute the empirical degree distribution $p_k$ from $G_0(V, 
	E)$\;
	Compute the size-biased distribution $q_k = \left(kp_k\right) / 
	\left(\sum_{k} k p_k\right)$\;
	Construct an appropriately-defined matrix $\bfM := (m_{kl})$ 
	such that the assortativity coefficient of the network 
	associated with 
	joint edge distribution $e_{kl} = q_k q_l + r^* \sigma_q^2 
	m_{kl}$ is $r^*$\;
	\While{$T > 0$}{
		Sample two edges $\{v_1, v_2\}, \{v_3, v_4\} \in E$ at 
		random\;
		Compute the degrees of the nodes at the ends of the two 
		sampled 
		edges:
		\\ \hspace{1cm}
		$i \leftarrow d_{v_1}$, $j \leftarrow 
		d_{v_2}$,
		$k \leftarrow d_{v_3}$, $l \leftarrow d_{v_4}$\;
		\eIf{$e_{ik}\, e_{jl} < e_{ij}\, e_{kl}$}
		{$p \leftarrow \left(e_{ik}\, e_{jl}\right) / \left(e_{ij}\, 
			e_{kl}\right)$\;}
		{$p \leftarrow 1$\;}
		Draw $U \sim \mathrm{Unif}(0, 1)$\; 
		\eIf{$U \leq p$}
		{Remove $\{v_1, v_2\}$ and $\{v_3, v_4\}$, append $\{v_1, 
		v_3\}$ 
			and $\{v_2, v_4\}$ \;}
		{Keep $\{v_1, v_2\}$ and $\{v_3, v_4\}$\;} 
		$T \leftarrow T - 1$ \;
	}
\end{algorithm}

A key step of the algorithm is to construct $\bfM$ with 
the following constraints: $\bfM$ is symmetric; both 
row and column sums are equal to $0$; 
$\sum_{kl} kl m_{kl} = 1$;
and the resulting $e_{kl}$'s are nonnegative.
The method in \citet{Newman2003mixing}, however, does not guarantee
the existence of $\bfM$ for any arbitrary~$r^*$, nor is it
directly applicable to directed networks.

\subsubsection{Directed Network Degree-Preserving Rewiring (DiDPR) 
Algorithm}
\label{sec:DiDPR}

Given an initial directed network $G_0(V, E)$ and four target 
assortativity coefficients (of different types), we propose a 
rewiring algorithm such that the four directed assortativity 
measures of the resulting network $G(V, E)$ will reach their 
corresponding targets simultaneously. 
The main idea of our algorithm is to characterize the network
$G(V, E)$ with the target
assortativity coefficients through $\eta_{ijkl}$, which 
is obtained by solving a convex optimization problem. 
During the entire procedure, the out- and in-degree distributions  
of $G_0(V, E)$ are preserved, and we refer to the proposed 
algorithm as a \emph{directed network degree-preserving rewiring} 
(DiDPR) algorithm.
We reuse $T$ to denote the number of iterative 
steps of rewiring. Similar to Newman's algorithm, we need $T$ to be 
sufficiently large to ensure the convergence of the proposed 
algorithm.

Recall the definition of $\eta_{ijkl}$ in Section~\ref{sec:dassort}.
Here we consider $\eta_{ijkl}$ as an extension of the joint edge 
distribution $e_{kl}$ in Algorithm~\ref{alg:newman}, and we will 
develop a 
reliable scheme to determine the four-dimensional matrix,
$\boldsymbol{\eta}:=\{\eta_{ijkl}\}$, 
which will then be used to characterize networks with the
given assortativity coefficients. According to the properties of 
$\eta_{ijkl}$ in Equations~\eqref{eq:restr1} 
and~\eqref{eq:restr2}, we notice that these linear relations are 
linear constraints in terms of $\eta_{ijkl}$, allowing us to 
convert the problem of finding $\boldsymbol{\eta}$ to 
a convex programming problem~\citep{Boyd2004convex}.

Suppose that
$r^*(a, b), a, b \in \{1, 2\}$, are the predetermined values of 
the four directed 
assortativity measures, and let $f(\cdot)$ be some convex function. 
According to the linear 
constraints on $\eta_{ijlk}$ given in Equations~\eqref{eq:restr1} 
and~\eqref{eq:restr2}, as well as the natural bounds of 
$\eta_{ijlk}$, we 
set up the following convex optimization problem to solve for 
appropriate $\boldsymbol{\eta}$, with the initial network 
$G_0(V, E)$:
\begin{align}
	\min_{\bm{\eta}} \qquad &f(\bm{\eta}),
	\nonumber\\ \textrm{s.t.} \qquad & -\eta_{ijkl} \le 0, 
	\label{opt}\\ &\sum_{k, l} \eta_{ijkl} = 
	\frac{i \nu_{ij}}{\sum_{i, j} i \nu_{ij}},
	\quad 
	\sum_{i, j} \eta_{ijkl} = 
	\frac{l \nu_{kl}}{\sum_{k, l} l \nu_{kl}},\nonumber\\
	&r\left(a, b \right) = r^* \left(a, b \right),\quad  a, b \in 
	\{1, 2\},\nonumber
\end{align}
where $r(a, b)$ for $a, b \in \{1, 2\}$, are functions of 
$\eta_{ijkl}$ and $\nu_{kl}$ by Equation~\eqref{eq:4dassort}. Since 
the proposed rewiring algorithm preserves out-degrees and 
in-degrees, the structure of $\bm{\nu} := (\nu_{kl})$ remains 
unchanged, allowing us to calculate all the values of $\nu_{kl}$ 
from the initial network $G_0(V, E)$. Specifically, we solve the 
convex optimization problem via the utility functions developed in 
\proglang{R} package \code{CVXR}~\citep{Rpkg:CVXR}, which is 
available on the \proglang{CRAN}. The \code{CVXR} package provides a 
user-friendly interface that allows users to formulate convex 
optimization problems in simple mathematical syntax, and utilizes 
some well developed algorithms, like the embedded conic solver 
(ECOS)~\citep{Domahidi2013ecos}, to solve the problems.

For the convex optimization problem defined above,
the four-dimensional structure $\bm{\eta}$ 
can be reduced to a matrix, where its elements are 
defined as non-negative variables, to fit the interface 
of the \code{CVXR} package. Details are presented in 
Appendix~\ref{Append:express}. Without loss of generality, we set 
the convex objective function as $f(\bm{\eta}) = 0$ to save 
computation powers.

Given the initial network $G_0(V, E)$ and the solved $\bm{\eta}$.
At each rewiring step, randomly select a pair of edges
$(v_1, v_2), (v_3, v_4) \in E$. Measure the out- and in 
degrees $i_1 = \dout_{v_1}$, $j_1 = \din_{v_1}$, 
$k_1 = \dout_{v_2}$, $l_1 = \din_{v_2}$, 
$i_2 = \dout_{v_3}$, $j_2 = \din_{v_3}$, 
$k_2 = \dout_{v_4}$ and $l_2 = \din_{v_4}$.
We then replace the selected $(v_1, v_2)$ and $(v_3, v_4)$ with 
$(v_1, 
v_4)$ and $(v_3, v_2)$ with probability 
\begin{equation}
	\label{eq:p}
	p = \begin{cases}
		\left(\eta_{i_1 j_1 k_2 l_2}\, \eta_{i_2 j_2 k_1 l_1}\right)/
		\left(\eta_{i_1 j_1 k_1 l_1}\, \eta_{i_2 j_2 k_2 
		l_2}\right), 
		&\qquad \text{if }
		\frac{\eta_{i_1 j_1 k_2 l_2}\, \eta_{i_2 j_2 k_1 
		l_1}}{\eta_{i_1 j_1 k_1 l_1}\, \eta_{i_2 j_2 k_2 l_2}} < 1
		;
		\\ 1, &\qquad \text{otherwise}.
	\end{cases}
\end{equation} 
We continue the rewiring in an iterative manner for $T$ times, and 
then obtain the resulting network $G(V, E)$, 
which is the output of the DiDPR algorithm. The pseudo codes of the 
DiDPR algorithm are given in Algorithm~\ref{alg:rewiring}.

This proposed algorithm has a few 
appealing properties. First, during the entire rewiring 
procedure, the out- and in-degrees of the nodes connected by the 
sampled edges remain unchanged regardless of the acceptance or 
rejection of the rewiring attempt, thus preserving the structure of 
$\bm{\nu}$. 
Next, the DiDPR algorithm is \emph{ergodic} over the 
collection of networks with given out- and in-degree sequences 
(denoted by $\mathcal{G}$), as any network in $\mathcal{G}$ can be 
reached within a finite number of rewiring steps. Lastly, the 
proposed algorithm satisfies the \emph{detailed balance} condition, 
i.e., for any 
two configurations $G_1, G_2 
\in \mathcal{G}$, it follows from Equation~\eqref{eq:p} that 
\[
\Pr(G_1) \Pr(G_1 \to G_2) = \Pr(G_2) \Pr(G_2 \to G_1),
\]
where $\Pr(G_1)$ denotes the probability of sampling configuration 
$G_1 \in \mathcal{G}$, and $\Pr(G_1 \to G_2)$ is the transition 
probability from $G_1$ to $G_2$. 

\begin{algorithm}[tbp]
	\caption{Pseudo codes for the DiDPR algorithm.}
	\label{alg:rewiring}
	\KwIn{Initial network $G_0(V, E)$, number of rewiring steps $T$, 
		target assortativity coefficients $r^*(a, b)$,
		$a, b \in \{1, 2\}$.}
	\KwOut{$G (V, E)$.}
	Apply the convex optimization algorithm to get $\bm{\eta}$\; 
	\While{$T > 0$}{
		Sample two directed edges $(v_1, v_2), (v_3, v_4) \in E$ at 
		random\;
		Compute the out- and in-degrees of the four nodes of the two 
		sampled edges:
		\\ \hspace{1cm}
		$i_1 \leftarrow \dout_{v_1}$, $j_1 \leftarrow 
		\din_{v_1}$, $k_1 \leftarrow \dout_{v_2}$, $l_1 
		\leftarrow \din_{v_2}$, 
		\\ \hspace{1cm} $i_2 \leftarrow \dout_{v_3}$, $j_2 
		\leftarrow \din_{v_3}$, $k_2 \leftarrow \dout_{v_4}$, $l_2 
		\leftarrow  \din_{v_4}$\;
		\eIf{$\eta_{i_1 j_1 k_2 l_2}\, \eta_{i_2 j_2 
				k_1 l_1} < \eta_{i_1 j_1 k_1 l_1}\, \eta_{i_2 
				j_2 k_2 l_2}$}
		{$p \leftarrow \left(\eta_{i_1 j_1 k_2 l_2}\, \eta_{i_2 j_2 
				k_1 l_1}\right) / \left(\eta_{i_1 j_1 k_1 l_1}\, 
				\eta_{i_2 
				j_2 k_2 l_2}\right)$\;}
		{$p \leftarrow 1$\;}
		Draw $U \sim \mathrm{Unif}(0, 1)$\;
		\eIf{$U \leq p$}
		{Remove $(v_1, v_2)$ and $(v_3, v_4)$, append $(v_1, v_4)$ 
			and $(v_3, v_2)$\;}
		{Keep $(v_1, v_2)$ and $(v_3, v_4)$\;}
		$T \leftarrow T - 1$\;
	}		
\end{algorithm}

\subsection{Directed Assortativity Coefficient Bounds}
\label{sec:range}

The four target assortativity coefficients, $r^*(a, b)$, 
$a, b \in \{1, 2\}$, are naturally bounded 
between $-1$ and $1$. 
However, based on the structure of $G_0(V, E)$, 
we cannot arbitrarily set four targets, $r^*(a, b)$, $a, b \in 
\{1, 2\}$, to any values within the range of $[-1, 1]$ while 
preserving $\bm{\nu}$, since certain combinations of the target 
values 
may not exist. 
In this section, we propose a method to determine the
bounds of $r^*(a, b)$ conditional on the initial configuration
$G_0(V, E)$.

Note that for a given $G_0(V,E)$, quantities like $q_k^{(a)}$, 
$\tq_l^{(b)}$, 
$\sigma_q^{(a)}$ and $\sigma_{\tq}^{(b)}$ are uniquely determined, 
and will remain unchanged throughout
the rewiring process. Then
by Equation~\eqref{eq:4dassort}, for 
$a,b \in \{1,2\}$, 
\[  \sum_{k, l} kl e^{\left(a, b\right)}_{kl} = 
g^{\left(a, b\right)} \bigl(r(a, b)\bigr) := \sigma_q^{(a)} 
\sigma_{\tq}^{(b)}\, r(a, b) + \sum_{k, l} kl q_k^{(a)} 
\tq_l^{(b)},
\]
which, confirms that
$\sum_{k, l} kl e^{\left(a, b\right)}_{kl}$
is a linear function of $r(a, b)$.

Next, we describe the procedure of finding the bounds of the four
assortativity coefficients. Without loss of generality, we take 
$r^*(1, 1)$ as our example. The upper and lower bounds of 
$r^*(1, 1)$ are
related to the structure of $G_0(V, E)$ which is 
characterized through its corresponding $\bm{\eta}$ 
and $\bm{\nu}$. We find the lower bound of $r^*(1, 1)$ by 
solving the following convex optimization problem:
\begin{align*}
	\min_{\bm{\eta}} \qquad &f^{(1,1)}(\bm{\eta}) = \sum_{i, k}ik 
	\left(\sum_{j, l} \eta_{ijkl}\right),
	\\ \textrm{s.t.} \qquad & -\eta_{ijkl} \le 0, 
	\\ &\sum_{k, l} \eta_{ijkl} = 
	\frac{i \nu_{ij}}{\sum_{i, j} i \nu_{ij}},
	\quad 
	\sum_{i, j} \eta_{ijkl} = 
	\frac{l \nu_{kl}}{\sum_{k, l} l \nu_{kl}}.
\end{align*}
Analogously, we obtain the upper bound of $r^{*}(1, 1)$ by 
solving the optimization problem with objective function $-f^{(1, 
	1)}(\bm{\eta})$, while keeping all of the
constraints unchanged. We denote the 
lower and upper bounds of $r^*(1, 1)$ as $r_{\rm L}^{*}(1, 1)$ 
and $r_{\rm U}^{*}(1, 1)$, respectively.

Now suppose that 
$U_{(1,1)}$ and $L_{(1,1)}$ are two predetermined values 
such that $r_{\rm L}^{*}(1, 1) \le 
L_{(1, 1)} \le U_{(1, 1)} \le r_{\rm U}^{*}(1, 1)$.
Then we determine the range of 
$r^{*}(1, 2)$, given the initial configuration $G_0(V, E)$ and
$L_{(1, 1)} \le r^*(1, 1) \le U_{(1, 1)}$. Here
the extra constraint of $r^*(1,1)\in [L_{(1,1)},U_{(1,1)}]$ 
further imposes restrictions on the possible values that 
$r^*(1,2)$ can take. 
The associated convex optimization problem for the lower bound of 
$r^*(1, 2)$ then becomes
\begin{align*}
	\min_{\bm{\eta}} \qquad &f^{(1,2)}(\bm{\eta}) = \sum_{i, l}il 
	\left(\sum_{j, k} \eta_{ijkl}\right),
	\\ \textrm{s.t.} \qquad & -\eta_{ijkl} \le 0, 
	\\ &\sum_{k, l} \eta_{ijkl} = 
	\frac{i \nu_{ij}}{\sum_{i, j} i \nu_{ij}},
	\quad 
	\sum_{i, j} \eta_{ijkl} = 
	\frac{l \nu_{kl}}{\sum_{k, l} l \nu_{kl}}
	\\ &g^{(1, 1)}\left(L_{(1, 1)}\right) \leq
	\sum_{i, k}ik \left(\sum_{j, l} \eta_{ijkl}\right) \leq 
	g^{(1, 1)}\left(U_{(1, 1)}\right).
\end{align*}
Similarly, the upper bound  of $r^*(1, 2)$ is obtained by 
solving the convex optimization problem with the same constraints 
but a different objective function 
$-f^{(1,2)}(\bm{\eta})$. We continue in this fashion until the 
bounds for all four assortativity coefficients are determined.

The proposed bound computation scheme provides a flexible framework 
so
that one may start with one arbitrary type of assortativity
coefficients, depending on the
information regarding the target network structure and the 
research problem of interest. Furthermore, the proposed scheme helps
determine whether the given target assortativity coefficients are 
attainable simultaneously, thus providing insights on their 
dependence structure. Since the DiDPR algorithm outlined in 
Algorithm~\ref{alg:rewiring}
tentatively costs a great deal of 
computation powers for mega scale networks, we suggest checking the 
attainability of the target assortativity coefficients before
applying the algorithm.

\section{Simulations}
\label{sec:sim}

We now investigate the performance of the DiDPR algorithm 
through simulation studies with
two widely used random network models, the
Erd\"{o}s--R\'{e}nyi (ER) model \citep{Erdos1959on, 
	Gilbert1959random} and Barab\'{a}si--Albert model 
\citep{Barabasi1999emergence}. The latter model is also known as 
linear preferential attachment (PA) network model in the 
literature.  We consider 
directed ER and PA models extended from their 
classical versions.

\subsection{ER Model}
\label{sec:ERmodel}

A directed ER random network, $\ER(n, p)$, is governed by 
two parameters: the number of nodes $n$ and the
probability of a directed edge from one node to 
another $p \in [0, 1]$. We consider an extension of the traditional 
ER random network model allowing self-loops.
In directed ER networks, all of the edges are generated 
independently, and
due to such simplicity, a 
variety of properties of ER networks have been investigated 
analytically; see for instance, \cite[Chapters 4 and 
5]{vdHofstad2017random}. 
Besides, ER random networks are often 
used as benchmark models in network analysis 
\cite[e.g.,][]{Bianconi2008local, Palla2015directed}.

We take $\ER(n, p)$ as our initial graph $G_0(V, E)$ with $\lvert 
V\rvert = n$. 
Since the directed edges in ER networks are generated independently, 
large-scale ER networks are not expected to present any patterns of 
assortative or disassortative mixing. Therefore, all of the values 
in the natural bound (i.e., $[-1, 1]$) are attainable for each of 
the assortativity coefficients marginally. 
Given one of the four 
assortativity coefficients, however, the 
values that the rest can take become restricted. Without 
loss of generality, we investigate the bounds of $r^*(1, 2)$, 
$r^*(2, 1)$ and $r^*(2, 2)$ conditional on the values of 
$r^*(1, 1) \in \{-0.9, -0.8, \ldots, 0.9\}$.

\begin{figure}[tbp]
	\centering
	\includegraphics[width = \textwidth]{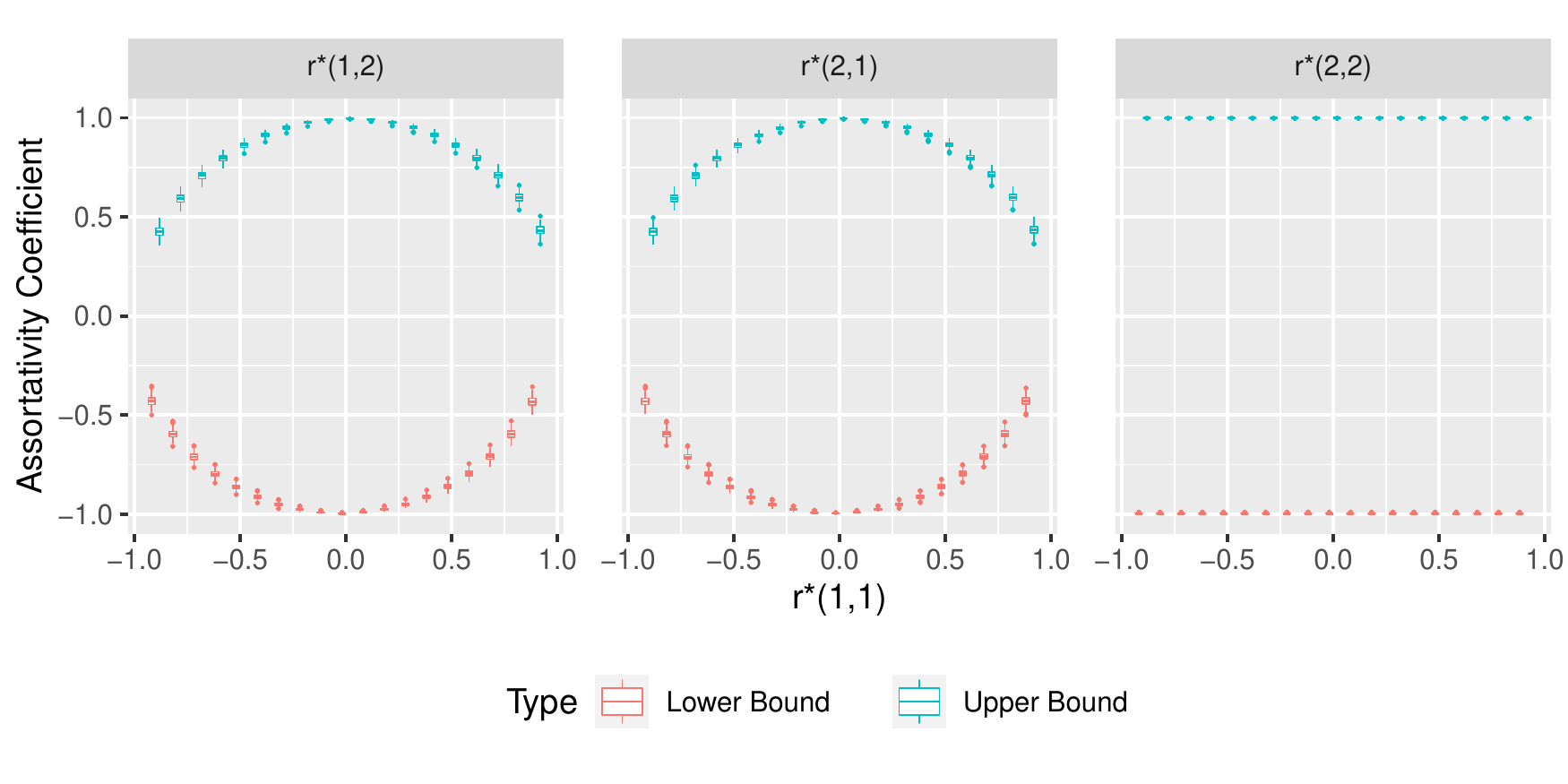}
	\caption{Side-by-side box plots of the upper and lower bounds of 
		$r^*(1, 2)$, $r^*(2, 1)$ and $r^*(2, 2)$ with given 
		values of $r^*(1, 1)$.}
	\label{fig:ER_bounds2}
\end{figure}

We generate $100$ independent ER networks with $n = 1000$ and 
$p = 0.1$ as initial graphs. For each initial graph, we solve the
corresponding convex optimization problem with
given $r^*(1, 1)$ values
to determine the upper and lower bounds of $r^*(1, 2)$,
$r^*(2, 1)$ and $r^*(2, 2)$. The results are presented via 
box plots in Figure~\ref{fig:ER_bounds2}. In the 
rightmost panel, the lower and upper bounds of 
$r^*(2, 2)$ respectively remain at $-1$ and $1$
regardless of changes in the values of $r^*(1, 1)$.
This suggests that the in-in degree correlation of ER networks is 
not 
affected by their out-out degree correlation, which agrees with the 
independence assumption made throughout the edge creation process. 
However, ranges of $r^*(1, 2)$ and $r^*(2, 1)$ 
are $[-1, 1]$ only when $r^*(1, 1)$ is close to $0$. Their bounds 
shrink symmetrically when the value of $r^*(1, 1)$ is deviated 
from $0$, since out-out 
assortativity coefficients with large magnitudes require a great
proportion of edges
linking source nodes with large (small) out-degree to target nodes
also with large (small) out-degree,
thus giving narrower bounds for 
out-in and in-out assortativity coefficients.

\begin{figure}[tbp]
	\centering
	\begin{subfigure}{0.9\textwidth}
		\includegraphics[width = \textwidth]{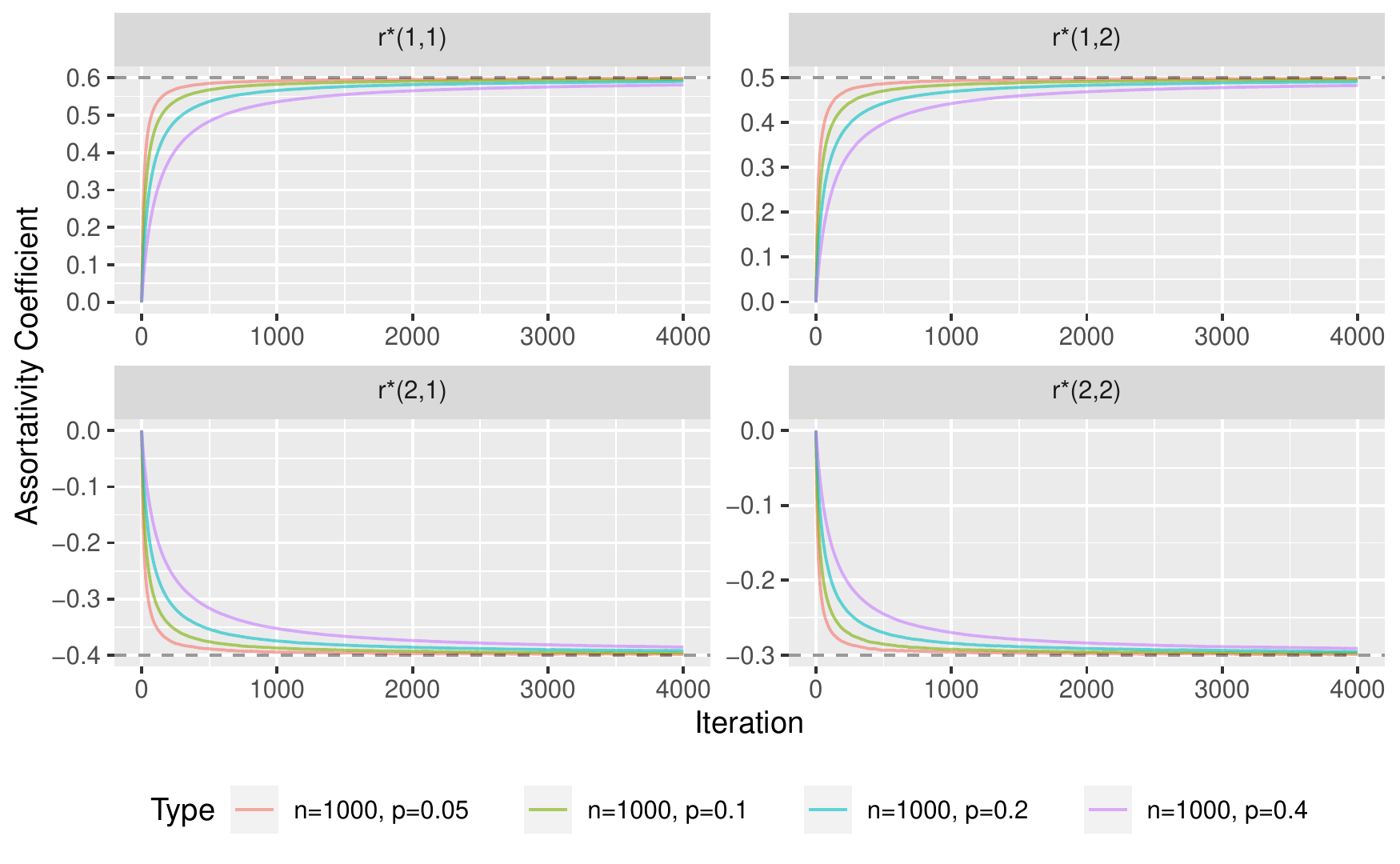}
	\end{subfigure}
	\begin{subfigure}{0.9\textwidth}
		\includegraphics[width = \textwidth]{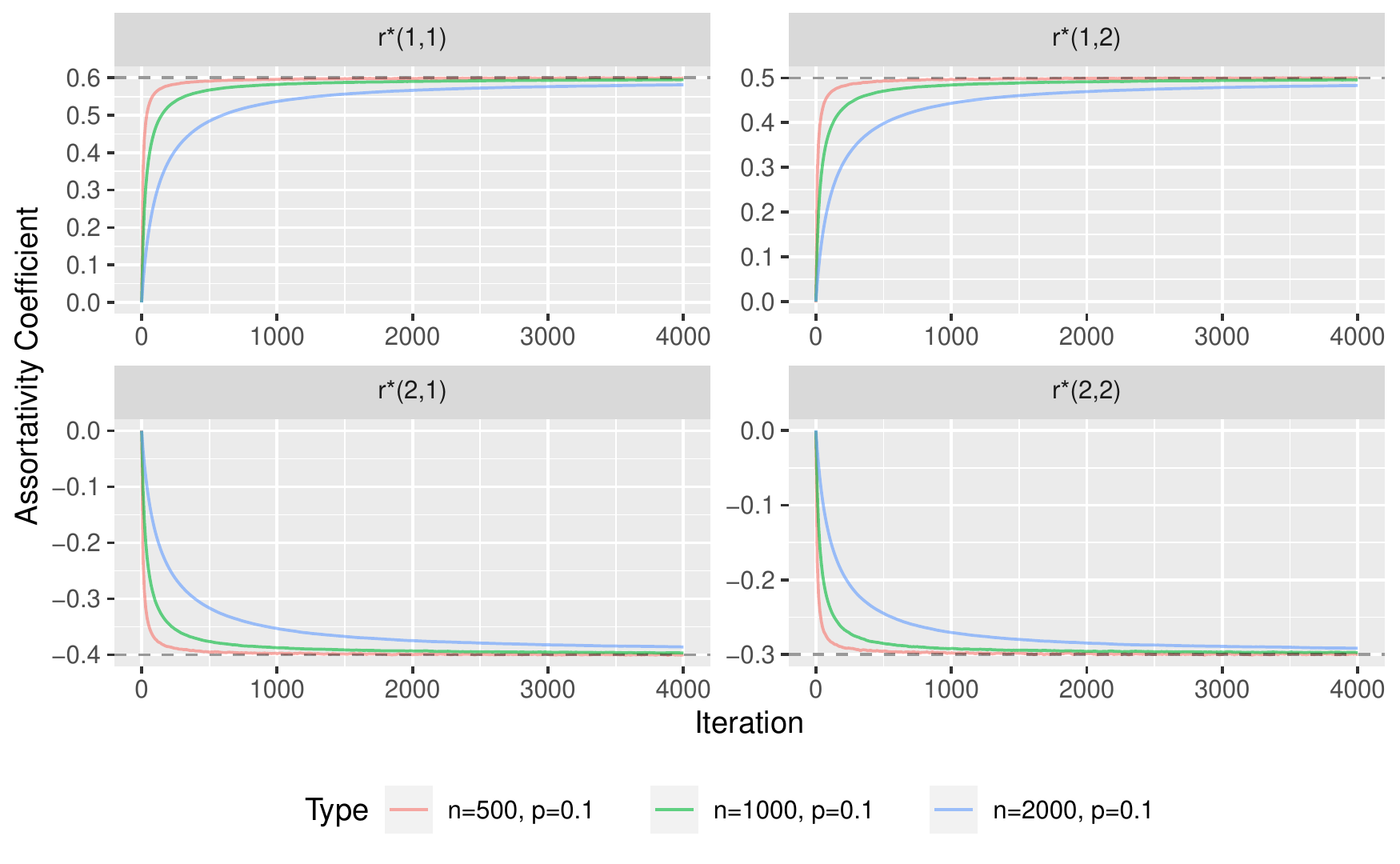}
	\end{subfigure}
	\caption{Average trace plots for the assortativity coefficients 
		of  directed ER networks. The parameters are set to $n = 
		1000$, 
		$p \in \{0.05, 0.1, 0.2, 0.4\}$ for the top four figures, 
		and 
		to $p = 0.1$, $n \in \{500, 1000, 2000\}$ for the bottom 
		four panels. The dashed grays lines represent the target 
		assortativity values.}
	\label{fig:ER_trace}
\end{figure}

Next, we conduct a sensitivity analysis to assess the performance 
of the DiDPR algorithm with respect to changes in $n$ and $p$. 
First, 
we fix $n = 1000$, and set $p \in \{0.05, 0.1, 0.2, 0.4\}$. The 
target assortativity values are given by $r^*(1, 1) 
= 0.6$, $r^*(1, 2) = 0.5$, $r^*(2, 1) = -0.4$ and 
$r^*(2, 2) = -0.3$, all of which are selected arbitrarily, and 
their attainability has been verified through our algorithm (from 
Section~\ref{sec:range}). For each combination of $n$ and $p$, we 
generate $100$ independent directed ER random networks, and present 
the average trace plots (of assortativity via rewiring) in Figure 
\ref{fig:ER_trace}, where each iteration contains $10^4$ rewiring 
steps. All trace plots in each panel start from $0$ as ER networks 
are not expected to show any pattern of assortative mixing.
For a fixed $n$ (in the top four panels of 
Figure~\ref{fig:ER_trace}), ER networks with smaller $p$ tend to 
arrive at the targets faster since only a small number 
of edges needs rewiring. On the other hand, we come up with the same 
conclusion according to the trace plots with fixed $p$ presented in 
the bottom four panels of Figure~\ref{fig:ER_trace}. Nonetheless, 
all of the trace plots confirm the success of the proposed algorithm.

\subsection{PA Model}
\label{sec:BAmodel}

\begin{figure}[tbp]
	\begin{center}
		\begin{tikzpicture}[scale=0.75]
			\draw[fill = gray!20] (-5, 0) ellipse (1.6 and 1) ;
			\draw (-4.2, 0) node[draw = black, circle, minimum 
			size =
			0.8cm, fill = blue!20] (i1)	{$i$} ;
			\draw (-3.5, -1.5) node[draw = black, circle, 
			minimum 
			size = 0.8cm] (u1)	{$u$} ;
			\draw[-latex, thick] (u1) -- (i1) ;
			\draw[fill = gray!20] (0, 0) ellipse (1.6 and 1) ;
			\draw (-0.8, 0) node[draw = black, circle, minimum 
			size =
			0.8cm, fill = blue!20] (i2)	{$i$} ;
			\draw (0.8, 0) node[draw = black, circle, minimum 
			size = 0.8cm, fill = blue!20] (j1)	{$j$} ;
			\draw[-latex, thick] (j1) -- (i2) ;
			\draw[fill = gray!20] (5, 0) ellipse (1.6 and 1) ;
			\draw (4.2, 0) node[draw = black, circle, minimum 
			size =
			0.8cm, fill = blue!20] (j2)	{$j$} ;
			\draw (3.5, -1.5) node[draw = black, circle, minimum 
			size = 0.8cm] (u2)	{$u$} ;
			\draw[-latex, thick] (j2) -- (u2) ;
		\end{tikzpicture}			
		\caption{Three edge-addition scenarios respectively 
			corresponding to $\alpha$, $\beta$ and $\gamma$ (from 
			left to	right).}
		\label{fig:PAex}
	\end{center}
\end{figure}
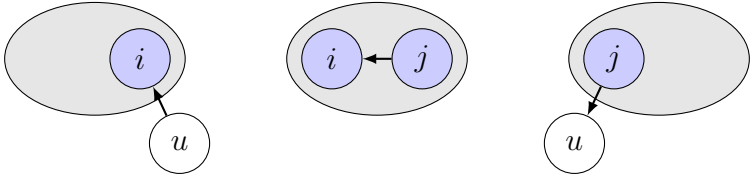

The PA model is a generative probabilistic model 
such that nodes with large degrees are more likely to attract
newcomers than those with small degrees
\citep[e.g.,][]{Barabasi1999emergence,Bollobas2003proceedings,
	Krapivsky2001degree, Krapivsky2001organization}. It is much more
realistic model for many real network data than the ER model.
We consider the directed PA (DPA) model 
given in~\citet{Bollobas2003proceedings}, which has
five parameters
$(\alpha, \beta, \gamma, \deltain, \deltaout)$ subject to
$\alpha + \beta + \gamma = 1$ as explained below.
Following \citet{Bollobas2003proceedings},
we assume there are three edge-creation
scenarios:
\begin{enumerate}
	\item With probability $0 \le \alpha \le 1$, a new edge is added 
	from a new node to an existing node following the PA rule.
	\item With probability $0 \le \beta \le 1$, a new edge is added 
	between two existing nodes following the PA rule.
	\item With probability $0 \le \gamma \le 1$, a new edge is added 
	from an existing node to a new node following the PA rule.
\end{enumerate}
A graphical illustration of this evolving process is given in 
Figure~\ref{fig:PAex}. The two offset parameters 
$\deltain, \deltaout > 0$ control the growth rate 
of in- and out-degrees, respectively \citep{Wang2021measuring}. 
The specific evolutionary rule of 
the model is given in Appendix~\ref{Append:PAgeneration}.

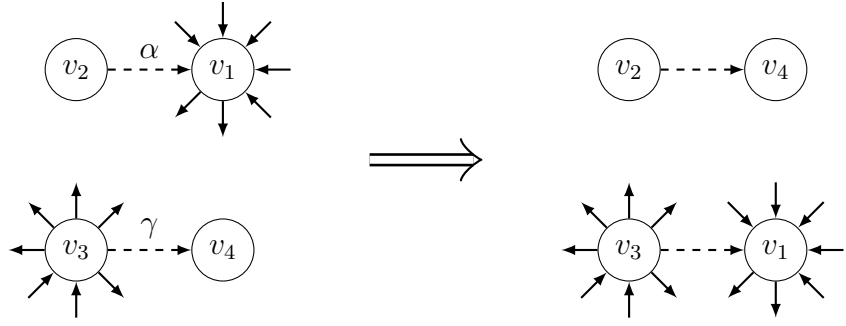
\begin{figure}[tbp]
	\begin{center}
		\begin{tikzpicture}[scale=0.75]
			\draw (-6.2, 1.6) node[draw = black, circle, minimum 
			size = 0.8cm]
			(v21) {$v_2$};
			\draw (-3.6, 1.6) node[draw = black, circle, minimum 
			size = 0.8cm]
			(v11) {$v_1$};
			\draw (-6.2, -1.6) node[draw = black, circle, minimum 
			size = 0.8cm]
			(v31) {$v_3$};
			\draw (-3.6, -1.6) node[draw = black, circle, minimum 
			size = 0.8cm]
			(v41) {$v_4$};
			\draw[-latex, thick, dashed] (v21) -- node [midway, 
			above] 
			{$\alpha$} (v11);
			\draw[-latex, thick] (-4.45, 2.45) -- (v11);
			\draw[-latex, thick] (-3.6, 2.8) -- (v11);
			\draw[-latex, thick] (-2.75, 2.45) -- (v11);
			\draw[-latex, thick] (-2.4, 1.6) -- (v11);
			\draw[-latex, thick] (-2.75, 0.75) -- (v11);
			\draw[-latex, thick] (v11) -- (-3.6, 0.4);
			\draw[-latex, thick] (v11) -- (-4.45, 0.75);
			
			\draw[-latex, thick, dashed] (v31) -- node [midway, 
			above] 
			{$\gamma$} (v41);
			\draw[-latex, thick] (v31) -- (-7.4, -1.6);
			\draw[-latex, thick] (v31) -- (-7.05, -0.75);
			\draw[-latex, thick] (v31) -- (-6.2, -0.4);
			\draw[-latex, thick] (v31) -- (-5.35, -0.75);
			\draw[-latex, thick] (v31) -- (-5.35, -2.45);
			\draw[-latex, thick] (-6.2, -2.8) -- (v31);
			\draw[-latex, thick] (-7.05, -2.45) -- (v31);
			
			\draw (3.6, 1.6) node[draw = black, circle, minimum size 
			= 0.8cm]
			(v22) {$v_2$};
			\draw (6.2, -1.6) node[draw = black, circle, minimum 
			size = 0.8cm]
			(v12) {$v_1$};
			\draw (3.6, -1.6) node[draw = black, circle, minimum 
			size = 0.8cm]
			(v32) {$v_3$};
			\draw (6.2, 1.6) node[draw = black, circle, minimum size 
			= 0.8cm]
			(v42) {$v_4$};
			\draw[-latex, thick, dashed] (v22) -- (v42);
			\draw[-latex, thick] (7.4, -1.6) -- (v12);
			\draw[-latex, thick] (7.05, -0.75) -- (v12);
			\draw[-latex, thick] (6.2, -0.4) -- (v12);
			\draw[-latex, thick] (5.35, -0.75) -- (v12);
			\draw[-latex, thick] (7.05, -2.45) -- (v12);
			\draw[-latex, thick] (v12) -- (6.2, -2.8);
			\draw[-latex, thick] (v12) -- (5.35, -2.45);
			
			\draw[-latex, thick, dashed] (v32) -- (v12);
			\draw[-latex, thick] (v32) -- (2.4, -1.6);
			\draw[-latex, thick] (v32) -- (2.75, -0.75);
			\draw[-latex, thick] (v32) -- (3.6, -0.4);
			\draw[-latex, thick] (v32) -- (4.45, -0.75);
			\draw[-latex, thick] (v32) -- (4.45, -2.45);
			\draw[-latex, thick] (3.6, -2.8) -- (v32);
			\draw[-latex, thick] (2.75, -2.45) -- (v32);
			
			\draw[line width=1pt, double distance=3pt, -{Classical 
			TikZ Rightarrow[length=3mm]}] (-1, 0) -- (1, 0); 
		\end{tikzpicture}
		\caption{Rewiring between $\alpha$-scenario and 
			$\gamma$-scenario edges, assuming $v_2>v_1$ and 
			$v_4>v_3$.}
		\label{fig:PArewire}
	\end{center}
\end{figure}

We provide a diagram in Figure~\ref{fig:PArewire} to explain how the 
rewiring process works for the DPA model. Suppose that two edges, 
$(v_2,v_1)$ and $(v_3,v_4)$, are sampled, assuming $(v_2,v_1)$ and 
$(v_3,v_4)$ are 
created under the $\alpha$- and $\gamma$-scenarios, respectively. 
According to the PA rule, node $v_1$ tends to have large 
in-degree, and $v_3$ tends to have large out-degree. However, the 
two 
nodes, $v_2$ and $v_4$, may have small in- and out-degrees since 
they 
are created at later stages of the network evolution. 
Then after a successful rewiring, we swap the edges to $(v_2, v_4)$ 
and 
$(v_3, v_1)$, increasing the assortativity coefficients.

\begin{figure}[tbp]
	\centering
	\includegraphics[width = \textwidth]{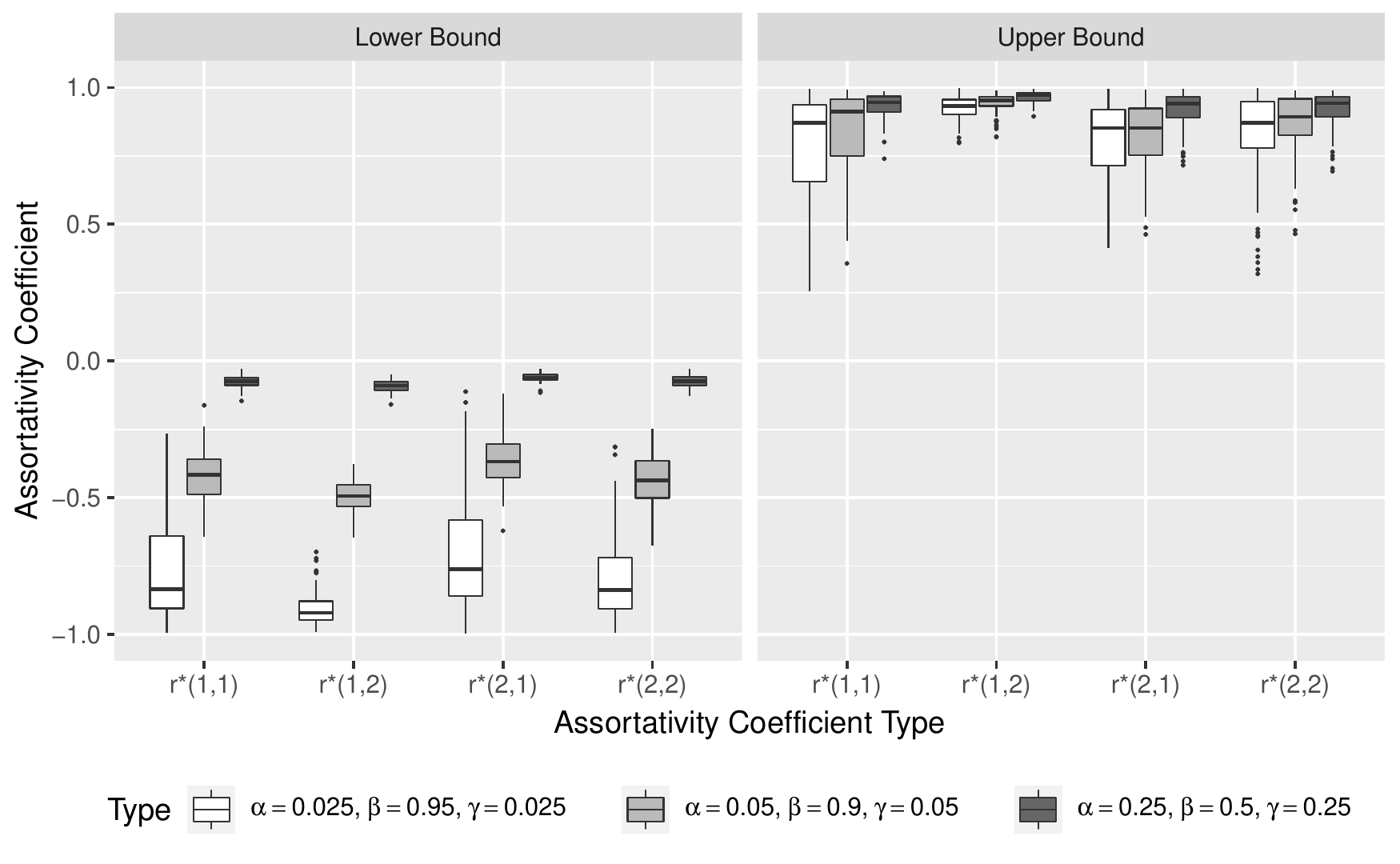}
	\caption{Side-by-side box plots of the upper and lower bounds 
		of the assortativity coefficients of DPA networks (of size 
		$10^5$) respectively 
		associated with parameters $\alpha = \gamma = 0.025, \beta = 
		0.95$; $\alpha = \gamma = 0.05, \beta = 0.9$; $\alpha = 
		\gamma = 
		0.25, \beta = 0.5$. For all of the generated PA networks, 
		$\deltaout$ and $\deltain$ are both set to $1$.}
	\label{fig:PA_bounds}
\end{figure}

Our simulation study starts with the investigation on the lower and 
upper
bounds of the four assortativity coefficients. We generate $100$ 
independent DPA networks of size $10^5$ with different sets of 
parameters, namely $\alpha = \gamma = 0.025$, 
$\beta = 0.95$, $\alpha = \gamma = 0.05$, 
$\beta = 0.9$ and $\alpha = \gamma = 0.25$, $\beta = 0.5$, while we 
set $\deltaout = \deltain = 1$ throughout the simulations.
Figure~\ref{fig:PA_bounds} presents the upper and lower bounds for
the four directed assortativity coefficients in the DPA networks with
three sets of parameters.  The range 
for large $\beta$ is wider than that for small $\beta$, 
with large (upper and lower) bound variations. For the 
DPA networks with $\alpha = \gamma = 0.05$, $\beta = 0.9$, the
upper and lower bounds of $r^*(1, 2)$, $r^*(2, 1)$ 
and $r^*(2, 2)$ given the value of
$r^*(1, 1) \in \{-0.1, 0, 0.1, 0.2, 0.3\}$
increase as $r^*(1, 1)$ increases.

\begin{figure}[tbp]
	\centering
	\includegraphics[width = \textwidth]{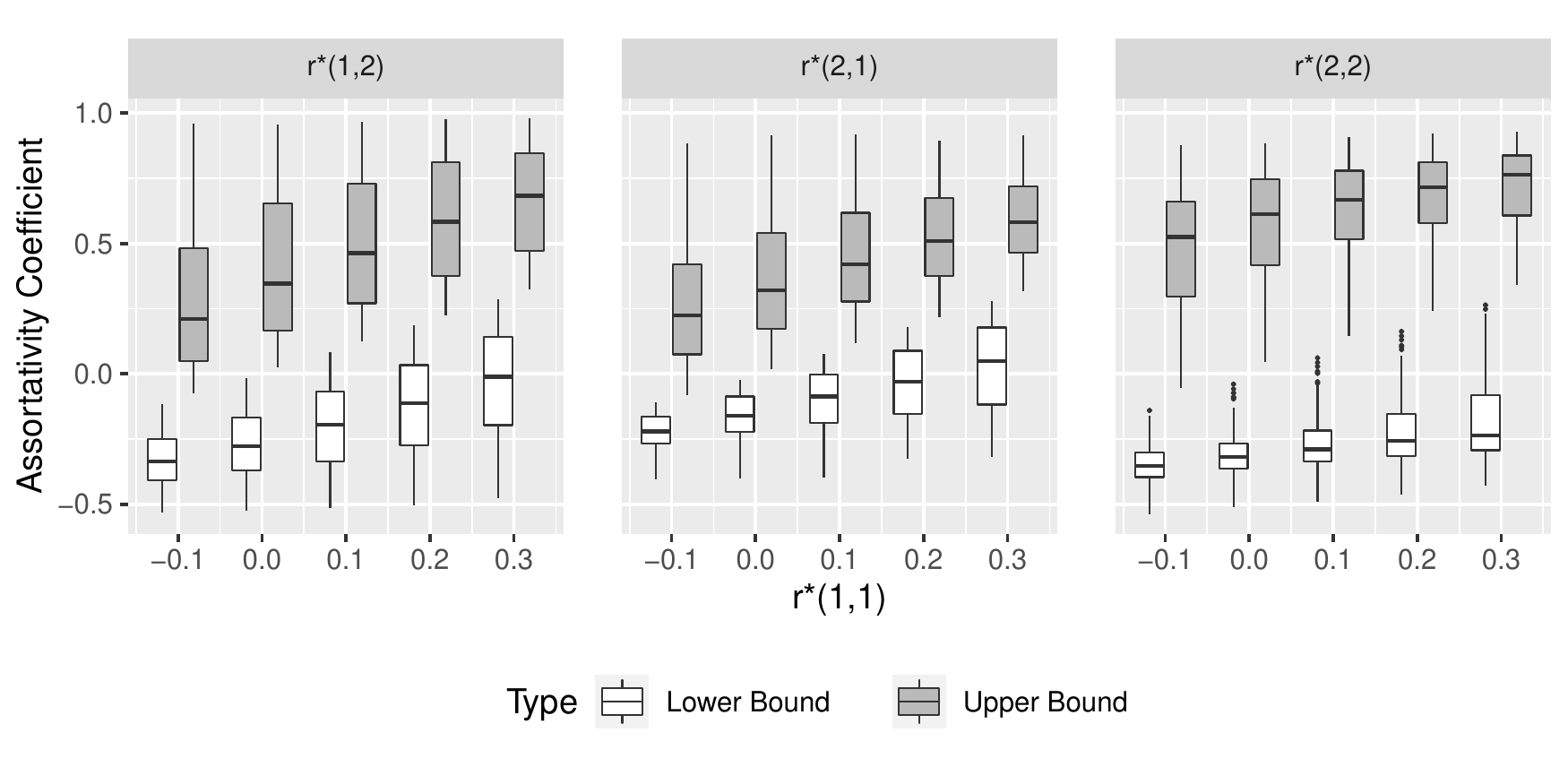}
	\caption{Side-by-side box plots of the upper and lower bounds of 
		$r^*(1, 2)$, $r^*(2, 1)$ and $r^*(2, 2)$ with given 
		values of $r^*(1, 1)$ of DPA networks of size $10^5$. The PA 
		parameters are set 
		to $\alpha = \gamma = 0.05, \beta = 0.9$, and $\deltaout = 
		\deltain = 1$.}
	\label{fig:PA_bounds2}
\end{figure}

We next assess the performance of the DiDPR algorithm
under DPA random networks with different
combinations of $(\alpha,\beta,\gamma)$ while holding
$\deltaout = \deltain = 1$. Moreover, we keep the four
assortativity targets as $r^*(1, 1) =  r^*(2, 1) = 0.1$ and 
$r^* (1, 2) = r^*(2, 2) = 0.15$. All of the assortativity 
targets are positive in response to the evolutionary feature of DPA 
networks, and the values are not too close to the extremes of 
lower/upper bounds, ensuring 
that they are achievable with high probabilities for different 
combinations of $(\alpha,\beta,\gamma)$ regardless of the 
structure of initial networks.

We also need to choose parameters in the simulation study carefully 
so that the 
message is articulated. Similar to the diagram in 
Figure~\ref{fig:PArewire},
we can draw other analogous rewiring diagrams when the sampled edges 
are from
$\alpha$-$\beta$, $\gamma$-$\beta$, $\alpha$-$\alpha$, 
$\beta$-$\beta$
and $\gamma$-$\gamma$ scenarios. After inspecting all combinations, 
we see that the $\alpha$-$\gamma$ combination provides the greatest 
amount of increase in the assortativity coefficients, supported by 
additional simulation experiments; see 
Appendix~\ref{Append:increase} for details. Hence, our simulation 
design has two different settings: (1) Fix $\alpha=\gamma$,
and vary values of $\beta$; (2) Fix $\beta$, and vary values of 
$\alpha\gamma$.
Under the first scenario, we maximize the chance of sampling the 
$\alpha$-$\gamma$ combination, and examine the impact of $\beta$ on 
the convergence of the assortativity coefficients.
In the second circumstances, by varying the product $\alpha\gamma$, 
we investigate 
whether a higher chance of sampling the $\alpha$-$\gamma$ 
combination gives 
faster convergence of the assortativity coefficients.

\begin{figure}[tbp]
	\centering
	\includegraphics[width = \textwidth]{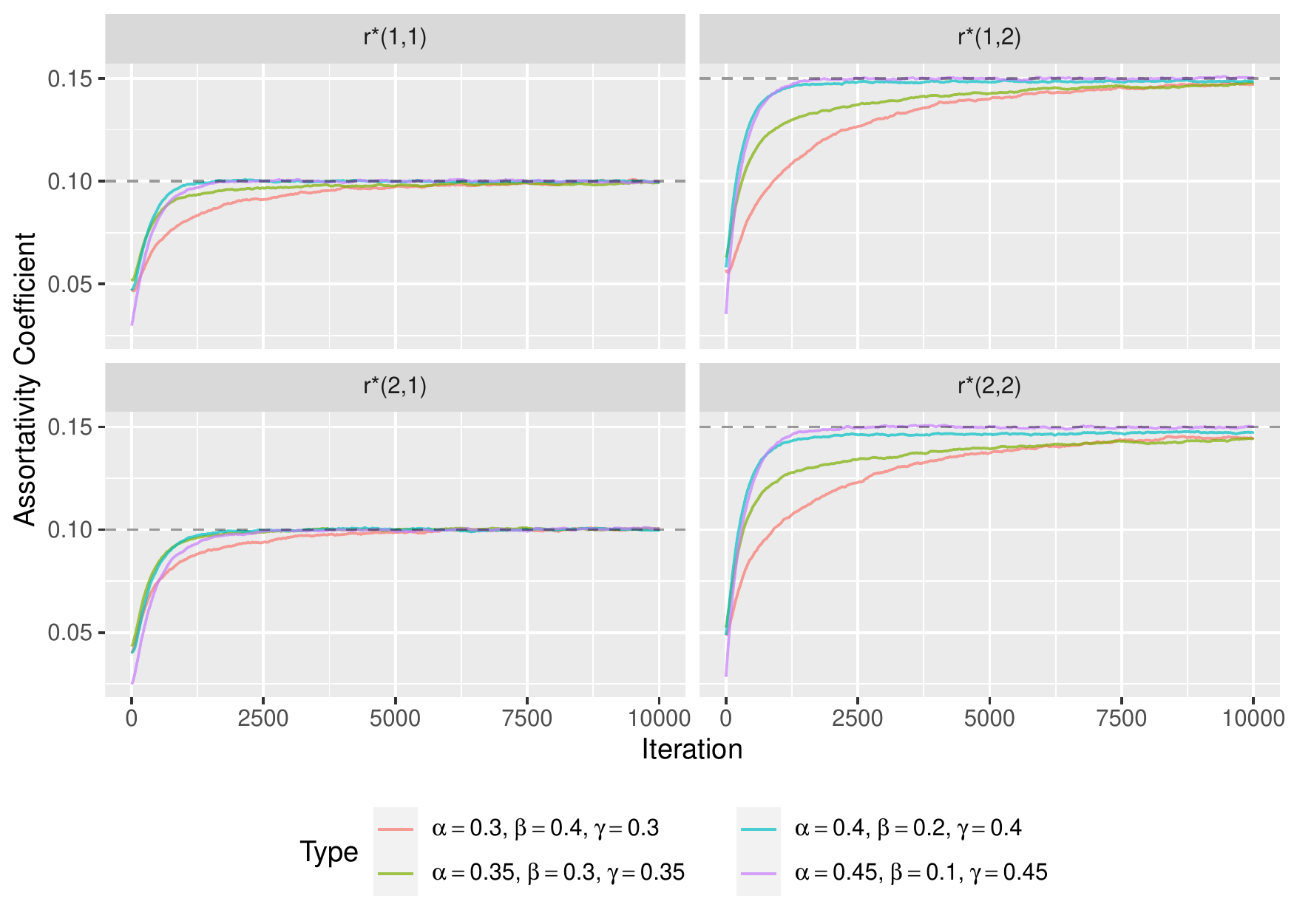}
	\caption{Average trace plots for four kinds of assortativity 
		coefficients of simulated DPA networks of size $100000$
		with $\alpha = \gamma$ and $\beta \in\{0.1, 0.2, 0.3, 
		0.4\}$.}
	\label{fig:alphaEqgamma}
\end{figure}

\paragraph{Fix $\alpha=\gamma$.}
Consider different 
values of $\beta \in \{0.1, 0.2, 0.3, 0.4\}$, and set the 
corresponding $\alpha = \gamma$ such that $\alpha + \beta + \gamma = 
1$. We do not allow $\beta$ to take large values in our simulations 
since otherwise 
the $\beta$-scenario will dominate the network evolution, decreasing 
the number of nodes 
created during the entire network growth. Similar to the previous 
study, we 
generate $100$ independent DPA networks, and collect the 
assortativity 
coefficient values every $10^3$ rewiring steps in order to improve 
the computational efficiency as well as for better 
graphical representation. The average trace plots are given 
in Figure~\ref{fig:alphaEqgamma}. 

The DiDPR algorithm shows rapid convergence of all four 
assortativity measures, and the assortativity coefficients 
reach their targets faster for smaller values of $\beta$.
When $\beta$ is small, we have a large amount of newly generated 
edges 
connecting existing nodes with new nodes. By the PA rule, 
existing nodes usually have larger out- and 
in-degrees than newcomers. Therefore, the initial graph are more 
likely 
to contain edges connecting large out-degree (in-degree) nodes with 
small in-degree (out-degree) nodes. While these edges are sampled, 
rewiring tentatively leads to
increases in assortativity values. Therefore, the proposed 
algorithm becomes effective and 
efficient. On the other hand, when $\beta$ is large, we expect many 
edges connecting existing nodes with large 
out- and in-degrees, and these edges are sampled with high 
probability. However, when two $\beta$-scenario edges are sampled, 
the improvement in assortativity will be
limited. Therefore, the assortativity coefficients in DPA networks 
with large $\beta$ require more time to attain the targets.

\begin{figure}[tbp]
	\centering
	\includegraphics[width = \textwidth]{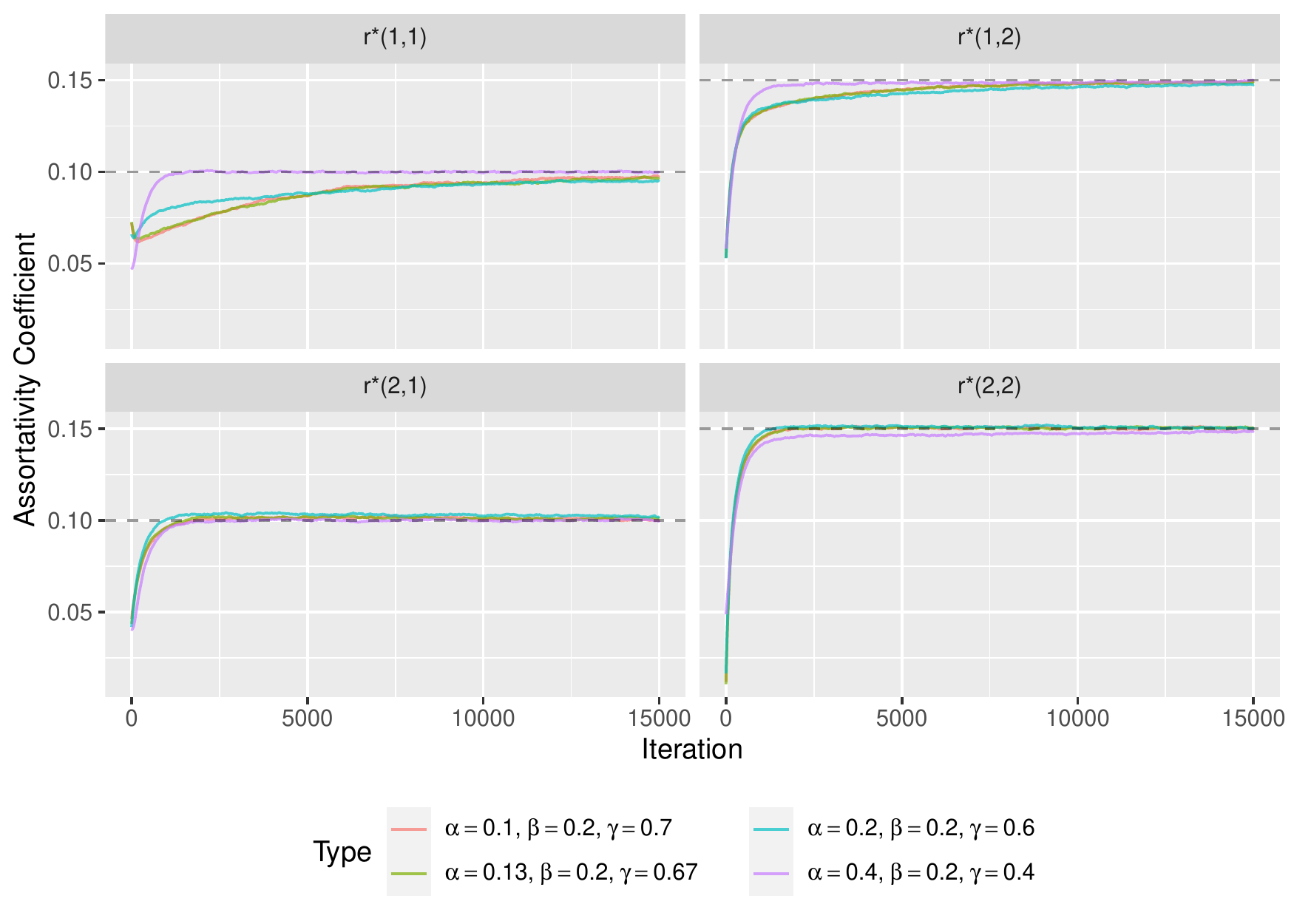}
	\caption{Average trace plots for the assortativity coefficients
		of simulated PA networks of size $10^5$ with $\beta = 0.2$ 
		and $\gamma / \alpha = c$ with $c \in \{1, 3, 5, 7\}$.}
	\label{fig:fixbeta}
\end{figure}

\paragraph{Fix $\beta$.}
We then fix $\beta = 0.2$, $\deltaout = 
\deltain = 1$ and consider different values of $\alpha$ and 
$\gamma$ such that $\gamma / \alpha = c$ for $c \in \{1, 3, 5, 7\}$. 
With the same target setting to $r^*(1, 1) =  r^*(2, 1) = 
0.1$ and $r^* (1, 2) = r^*(2, 2) = 0.15$, we generate $100$ 
independent DPA networks, and present the average trace plots in 
Figure~\ref{fig:fixbeta}. Once again, the DiDPR algorithm gives fast
convergence for all four assortativity coefficients. For 
$r^*(1,1)$ and $r^*(1, 2)$, they reach their corresponding
targets faster when $c$ is small. 
Since $c=1$ maximizes the value of $\alpha\gamma$ for a fixed 
$\beta$,
the fast convergence for small $c$ coincides with the earlier remark 
that 
the $\alpha$-$\gamma$ sampling combination gives the largest amount 
of improvement in the assortativity coefficients after each rewiring 
attempt.
The average trace plots for $r^*(2,1)$ and $r^*(2, 2)$, however,
do not display huge discrepancies in the convergence rate under 
different
parameter choices.


\section{Applications to Social Networks}
\label{sec:appl}

We now apply the proposed algorithm to a Facebook wall post network
with data available at KONECT
\citep[\url{http://konect.cc/networks/facebook-wosn-wall/},][]
{Kunegis2013konect}. We fit a DPA model to the selected network, and 
estimate the parameters via an {\em extreme value} (EV) method. The 
fitted model well captures the features of out- and in-degree 
distributions of the network data, but fails to characterize the 
assortativity structure accurately. By applying the DiDPR algorithm, 
we see that four assortativity coefficients of the fitted model are 
close to the counterparts of the selected network.

Nodes in the Facebook wall post network correspond to
Facebook users, and each directed edge $(i,j)$ represents 
the event that user~$i$ writes a post on the wall of user~$j$.
We only use the network formed by the data from 2007-07-01 to
2007-11-30 due to the observation in \citet{Wang2021common} 
that the network growth pattern in this period is
more stable than early time periods. The selection has led a network
with $16,549$ nodes and $147,063$ edges.

We start by fitting a DPA model to the network.
Different from the likelihood based method in \citet{Wan2017fitting},
the EV estimation method given in
\citet{Wan2020extreme}
only focuses on the distribution for large in- and out-degrees as
opposed to the entire network evolution history. Since our DiDPR
algorithm does not change the degree distributions, we use
the EV method to fit the DPA model in the first place.
Overall, the EV method considers a reparametrization of the DPA model
with unknown parameters
$(\alpha,\beta,\iota_1,\iota_2)$, where $\iota_1,\iota_2$ are the
marginal tail indices for the out- and in-degree distributions,
respectively. By \citet{Bollobas2003proceedings}, $(\iota_1,\iota_2)$
are functions of $(\alpha,\beta,\deltain,\deltaout)$:
\[\iota_1 = \frac{1 + \deltaout(\alpha + \gamma)}{\beta + \gamma} 
\qquad \text{and} \qquad \iota_2 = \frac{1 + \deltain(\alpha + 
	\gamma)}{\alpha + \beta}.\]

To implement the EV method, we first estimate $\beta$ by
$\hat\beta:= 1-\lvert V\rvert/\lvert E\rvert$.
Then to obtain the marginal out- and in-degree 
tail estimates, $\hat\iota_{1}$ and $\hat\iota_{2}$, we consult the 
\emph{minimum distance method} proposed in \citet{clauset2009power}, 
which is implemented in the \proglang{R} package 
\code{poweRlaw}~\citep{Rpkg:poweRlaw}.
Set $\hat{a}:= \hat\iota_2 / \hat\iota_1$, then by applying the
power transformation 
$(\dout_v,\din_v) \mapsto \left(\dout_v, (\din_v)^{\hat{a}}\right)$, 
we see that the transformed pair will have the same marginal tail 
index.
Next, we apply the polar transformation under the $L_1$-norm to 
obtain
\begin{equation*}
	\left(\dout_v, \din_v\right) \mapsto \left(
	\dout_v + \left(\din_v\right)^{\hat a}, 
	\frac{\left(\din_v\right)^{\hat a}}{\dout_v + 
	\left(\din_v\right)^{\hat a}}
	\right) := \left(R_v, \theta_v\right), \qquad v\in V.
\end{equation*}
Following the methodology in \citet{Wan2020extreme}, 
we estimate $\hat{\alpha}$ from the empirical distribution of 
$\theta_v$ for which $R_v > c$, and $c$ is typically chosen as 
the $(n_{\, \rm{tail}} + 1)$-th largest value of $\{R_v:v\in V\}$.
Using $n_{\, \rm{tail}} = 200$, we have $\hat{\alpha} = 0.008$, 
$\hat{\beta} = 0.887$, $\hat{\gamma} = 0.105$, $\hat{\delta}_{\, \rm 
	out} = 10.432$ and $\hat{\delta}_{\, \rm in} = 6.078$.

\begin{figure}[tbp]
	\includegraphics[width=\linewidth]{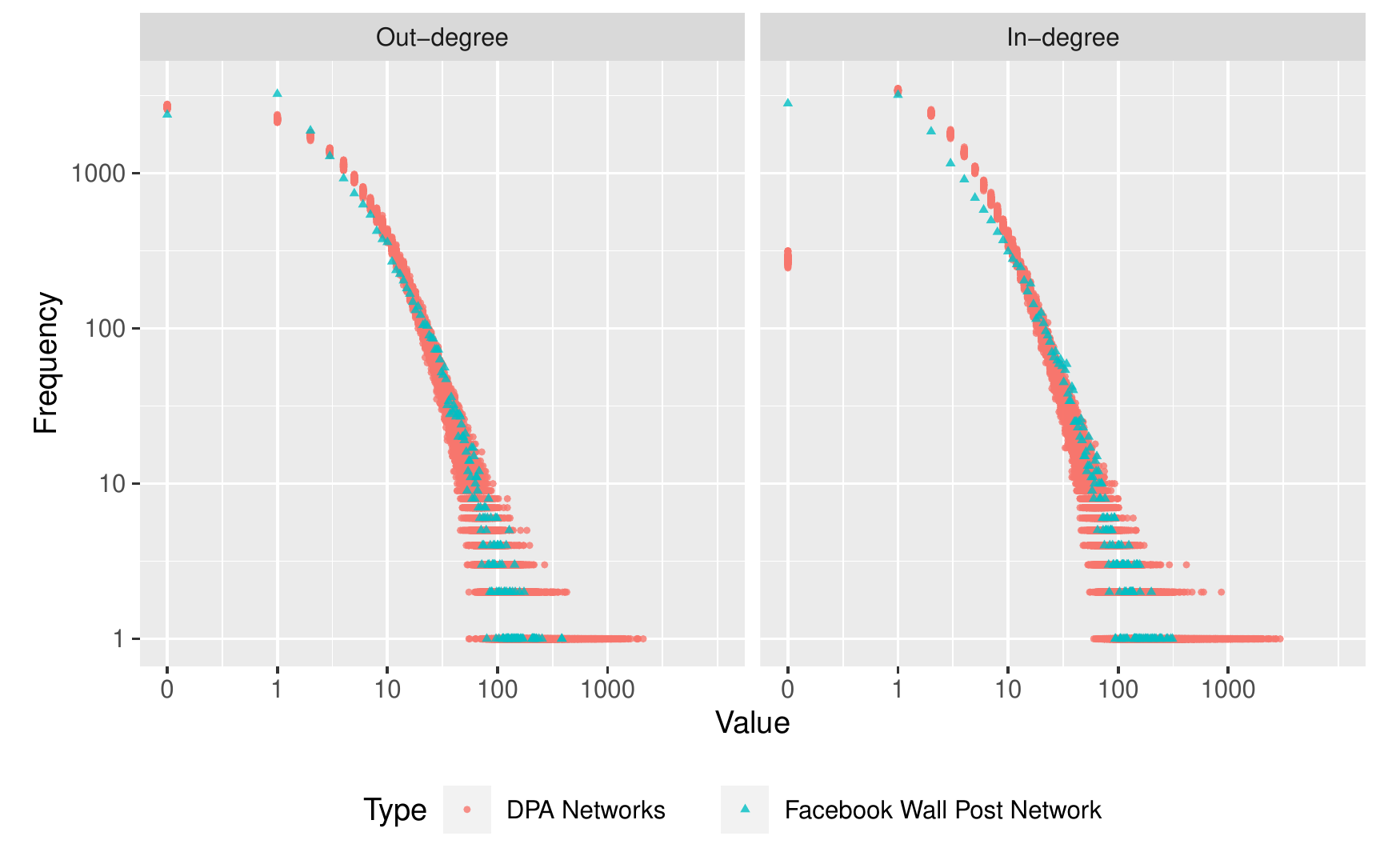}
	\caption{Empirical out- and in-degree distributions of the 
		selected sub-network and those from the $100$ independently 
		generated DPA networks with estimated parameters.}
	\label{fig:degree}
\end{figure}

We then generate $100$ independent DPA 
networks of size $147,063$ with the estimated parameters, and 
overlay the marginal out- and in-degree distributions of the 
simulated
networks and their
empirical counterparts from the selected sub-network; see
Figure~\ref{fig:degree}.  Most of the empirical out- and
in-degree distributions of the real data fall within or close to the 
ranges formed by the simulated networks, except for in-degree~$0$. 
Such discrepancy is due to the fact that there exists a certain 
number of users
who keep posting on others' Facebook walls, but have not received 
any posts during 
the observational period. Hence, the fitted DPA model is able to 
capture the degree
distribution in the given network, which provides the foundation for 
the implementation
of the DiDPR algorithm.

\begin{figure}[tbp]
	\centering
	\includegraphics[width = \textwidth]{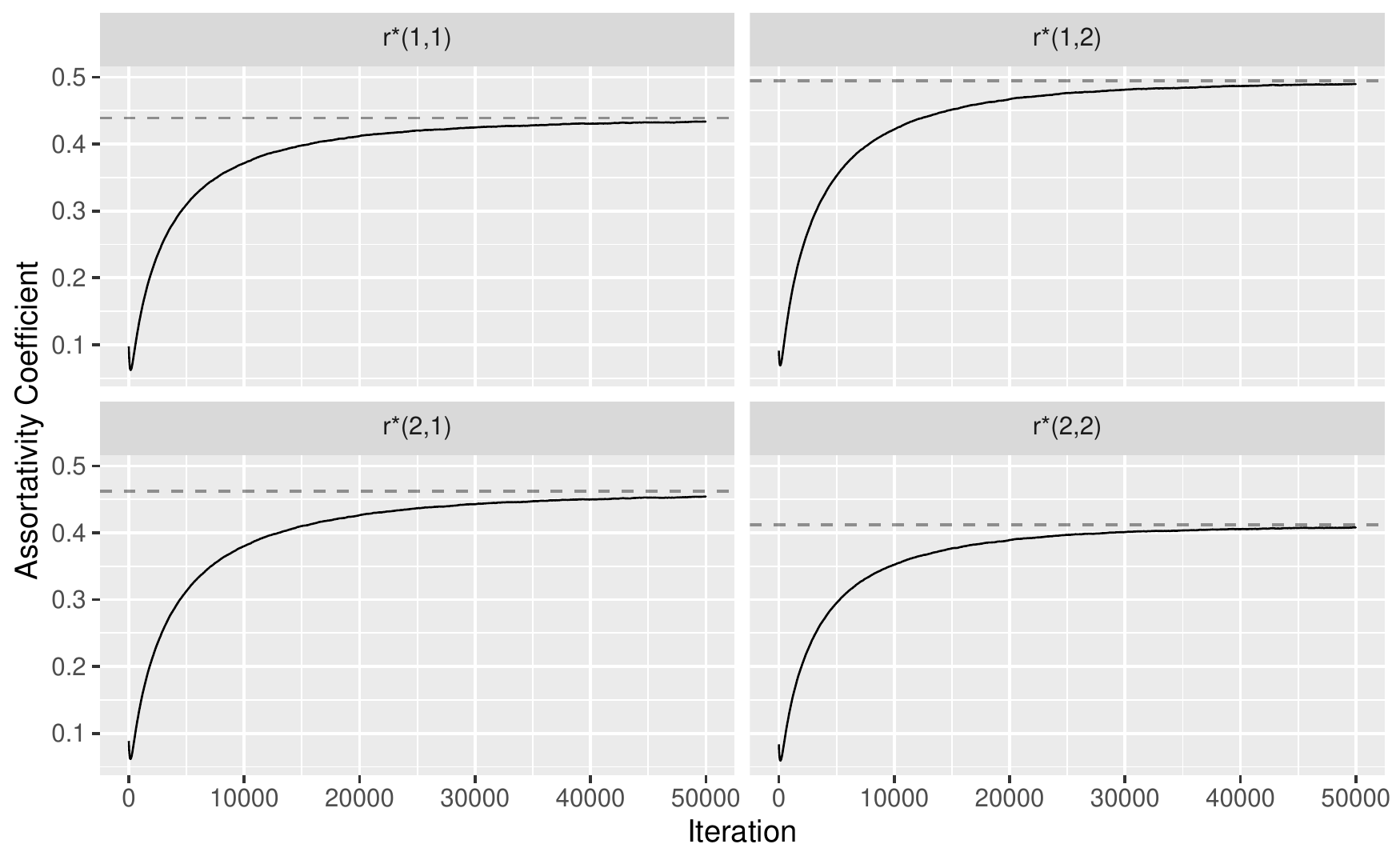}
	\caption{Average trace plots for the assortativity coefficients
		of simulated DPA networks.}
	\label{fig:facebook}
\end{figure}

Looking at the averages for the four assortativity values of the 
simulated networks, we have $r(1, 1) = 0.10$, $r(1, 
2) = 0.09$, $r(2, 1) = 0.09$, and $r(2, 2) = 0.08$, all of which are 
lower than their counterparts in the empirical
network, i.e., $r(1, 1) = 0.44$, $r(1, 2) = 0.49$, $r(2, 1) = 0.46$, 
and $r(2, 2) = 0.41$. Hence, we proceed by first using the DPA 
network with estimated 
parameters as initial configuration, then applying the DiDPR 
algorithm to 
correct the assortativity levels of the network, keeping the well 
fitted degree distributions unchanged. Figure~\ref{fig:facebook} 
shows the average trace plots of the assortativity coefficients 
based on the $100$ simulated DPA networks, where the assortativity 
values of each kind are updated every $10^3$ rewiring steps. 
Figure~\ref{fig:facebook} 
confirms that after rewiring, all of the assortativity coefficients 
are 
close to their counterparts observed from the selected sub-network, 
thus
filling up the discrepancy in the simple DPA model.

\section{Discussion}
\label{sec:dis}

The proposed DiDPR algorithm is efficient and effective in
generating directed networks with four pre-determined directed
assortativity coefficients. 
The fundamental step of the algorithm is to construct a directed
network achieving the given assortativity coefficients,
which is done by solving a convex optimization problem. This 
procedure complements a crucial missing component in Newman's 
rewiring
algorithm for undirected networks. With minor modifications, our
method can
identity the bounds of the assortativity coefficients by capturing 
the dependence structure among them. The proposed algorithm corrects 
all of the assortativity coefficient values simultaneously through 
rewiring process while preserving the original out- and in-degree 
distributions. The 
effectiveness of the algorithm is reflected through
simulation studies as well as an application to Facebook wall post 
data.

The proposed DiDPR algorithm
can be employed to adjust the assortativity coefficients 
defined similarly to Pearson's correlation. For instance, 
\citet{Van2015degree} proposed a rank-based assortativity 
coefficient analogous to Spearman's $\rho$ for undirected networks. 
We can define four (Spearman's $\rho$) rank-based assortativity 
coefficients for directed networks, where all the degree terms in 
Equation~\eqref{eq:4dassort} are replaced with the corresponding 
ranks. Mid-rank can be used to in presence of ties.
The same idea can be carried over to 
construct $\bm{\eta}$ for given assortativity targets. The rewiring 
procedure in Algorithm~\ref{alg:rewiring} remains unchanged for 
preserving out- and in-degree distributions. Therefore, the DiDPR 
algorithm can be adapted to directed assortativity 
coefficients defined with Spearman's $\rho$ straightforwardly, and
potentially to other directed assortativity defined with 
nonparametric
dependence measures such as Kendall's $\tau$.

A future direction of interest is to extend the algorithm to 
weighted, 
directed networks while preserving the strength distributions
throughout rewiring. This generalization is simple for 
integer-valued edge weights, as it can be decomposed into multiple 
unit-weighted edges. Preserving node strengths of continuous type,
however, remains challenging, especially when swapping two edges with
different weights.



\begin{thebibliography}{34}
\newcommand{\enquote}[1]{``#1''}
\expandafter\ifx\csname natexlab\endcsname\relax\def\natexlab#1{#1}\fi

\bibitem[{Barab{\'a}si and Albert(1999)}]{Barabasi1999emergence}
Barab{\'a}si, A.-L. and Albert, R. (1999), \enquote{Emergence of Scaling in
  Random Networks,} \textit{Science}, 286, 509--512.

\bibitem[{Bertotti and Modanese(2019)}]{Bertotti2019configuration}
Bertotti, M.~L. and Modanese, G. (2019), \enquote{The Configuration Model for
  {B}ar\'{a}basi-{A}lbert Networks,} \textit{Applied Network Science}, 4, 32.

\bibitem[{Bianconi et~al.(2008)Bianconi, Gulbahce, and
  Motter}]{Bianconi2008local}
Bianconi, G., Gulbahce, N., and Motter, A.~E. (2008), \enquote{Local Structure
  of Directed Networks,} \textit{Physical Review Letters}, 100, 11.

\bibitem[{Bollob\'{a}s et~al.(2003)Bollob\'{a}s, Borgs, , Chayes, and
  Riordan}]{Bollobas2003proceedings}
Bollob\'{a}s, B., Borgs, C., , Chayes, J., and Riordan, O. (2003),
  \enquote{Directed Scale-Free Graphs,} in \textit{SODA '03: Proceedings of the
  Fourteenth Annual ACM-SIAM Symposium on Discrete Algorithms}, Philadelphia,
  PA, USA: SIAM, pp. 132--139.

\bibitem[{Boyd and Vandenberghe(2004)}]{Boyd2004convex}
Boyd, S. and Vandenberghe, L. (2004), \textit{Convex Optimization}, Cambridge,
  U.K.: Cambridge University Press.

\bibitem[{Chang et~al.(2007)Chang, Su, Zhou, and He}]{Chang2007assortativity}
Chang, H., Su, B.-B., Zhou, Y.-P., and He, D.-R. (2007), \enquote{Assortativity
  and Act Degree Distribution of Some Collaboration Networks,} \textit{Physica
  A: Statistical Mechanics and Its Applications}, 383, 687--702.

\bibitem[{Clauset et~al.(2009)Clauset, Shalizi, and Newman}]{clauset2009power}
Clauset, A., Shalizi, C.~R., and Newman, M. E.~J. (2009), \enquote{Power-law
  distributions in empirical data,} \textit{SIAM Review}, 51, 661--703.

\bibitem[{Domahidi et~al.(2013)Domahidi, Chu, and Boyd}]{Domahidi2013ecos}
Domahidi, A., Chu, E., and Boyd, S. (2013), \enquote{{ECOS}: {A}n {SOCP} Solver
  for Embedded Systems,} in \textit{2013 European Cotrol Conference (ECC)},
  Piscataway, NJ, USA: IEEE, pp. 3071--3076.

\bibitem[{Erd\"{o}s and R\'{e}nyi(1959)}]{Erdos1959on}
Erd\"{o}s, P. and R\'{e}nyi, A. (1959), \enquote{On Random Graphs {I},}
  \textit{Publicationes Mathematicae Debrecen}, 6, 290--297.

\bibitem[{Foster et~al.(2010)Foster, Foster, Grassberger, and
  Paczuski}]{Foster2010edge}
Foster, J.~G., Foster, D.~V., Grassberger, P., and Paczuski, M. (2010),
  \enquote{Edge Direction and the Structure of Networks,} \textit{Proceedings
  of the National Academy of Sciences of the United States of America}, 107,
  10815--10820.

\bibitem[{Fu et~al.(2020)Fu, Narasimhan, and Boyd}]{Rpkg:CVXR}
Fu, A., Narasimhan, B., and Boyd, S. (2020), \enquote{{CVXR}: {A}n {R} Package
  for Disciplined Convex Optimization,} \textit{Journal of Statistical
  Software}, 94, 1--34.

\bibitem[{Gilbert(1959)}]{Gilbert1959random}
Gilbert, E.~N. (1959), \enquote{Random Graphs,} \textit{Annals of Mathematical
  Statistics}, 30, 1141--1144.

\bibitem[{Gillespie(2015)}]{Rpkg:poweRlaw}
Gillespie, C.~S. (2015), \enquote{Fitting Heavy Tailed Distributions: {T}he
  {poweRlaw} Package,} \textit{Journal of Statistical Software}, 64, 1--16.

\bibitem[{Holme and Zhao(2007)}]{Holme2007exploring}
Holme, P. and Zhao, J. (2007), \enquote{Exploring the Assortativity-Clustering
  Space of a Network's Degree Sequence,} \textit{Physical Review E}, 75,
  046111.

\bibitem[{Kashyap and Ambika(2017)}]{Kashyap2017mechanism}
Kashyap, G. and Ambika, G. (2017), \enquote{Mechanisms for Tuning Clustering
  and Degree-Correlations In Directed Networks,} \textit{Journal of Complex
  Networks}, 6, 767--787.

\bibitem[{Krapivsky and Redner(2001)}]{Krapivsky2001organization}
Krapivsky, P.~L. and Redner, S. (2001), \enquote{Organization of Growing Random
  Networks,} \textit{Physical Review E}, 63, 066123.

\bibitem[{Krapivsky et~al.(2001)Krapivsky, Rodgers, and
  Redner}]{Krapivsky2001degree}
Krapivsky, P.~L., Rodgers, G.~J., and Redner, S. (2001), \enquote{Degree
  Distributions of Growing Networks,} \textit{Physical Review Letters}, 86,
  5401--5404.

\bibitem[{Kunegis(2013)}]{Kunegis2013konect}
Kunegis, J. (2013), \enquote{{KONECT}: {T}he {K}oblenz Network Collection,} in
  \textit{WWW ’13 Companion: Proceedings of the 22nd International Conference
  on World Wide Web}, eds. Schwabe, D., Almeida, V., and Glaser, H., New York,
  NY, USA: Association for Computing Machinery, pp. 1343--1350.

\bibitem[{Leung and Chau(2007)}]{Leung2007weighted}
Leung, C.~C. and Chau, H.~F. (2007), \enquote{Weighted Assortative and
  Disassortative Networks Model,} \textit{Physica A: Statistical Mechanics and
  Its Applications}, 378, 591--602.

\bibitem[{Litvak and van~der Hofstad(2013)}]{Litvak2013uncovering}
Litvak, N. and van~der Hofstad, R. (2013), \enquote{Uncovering Disassortativity
  in Large Scale-Free Networks,} \textit{Physical Review E}, 87, 022801.

\bibitem[{Newman(2002)}]{Newman2002assortative}
Newman, M. E.~J. (2002), \enquote{Assortative Mixing in Networks,}
  \textit{Physical Review Letters}, 89, 208701.

\bibitem[{Newman(2003)}]{Newman2003mixing}
--- (2003), \enquote{Mixing Patterns in Networks,} \textit{Physical Review E},
  67, 026126.

\bibitem[{Noldus and van Mieghem(2015)}]{Noldus2015assortativity}
Noldus, R. and van Mieghem, P. (2015), \enquote{Assortativity in Complex
  Networks,} \textit{Journal of Complex Networks}, 3, 507--542.

\bibitem[{Palla et~al.(2015)Palla, Farkas, Pollner, Der\'{e}nyi, and
  Vicsek}]{Palla2015directed}
Palla, G., Farkas, I.~J., Pollner, P., Der\'{e}nyi, I., and Vicsek, T. (2015),
  \enquote{Directed Network Modules,} \textit{New Journal of Physics}, 9, 186.

\bibitem[{Piraveenan et~al.(2012)Piraveenan, Prokopenko, and
  Zomaya}]{Piraveenan2012assortative}
Piraveenan, M., Prokopenko, M., and Zomaya, A. (2012), \enquote{Assortative
  Mixing in Directed Biological Networks,} \textit{IEEE/ACM Transactions on
  Computational Biology and Bioinformatics}, 9, 66--78.

\bibitem[{Uribe-Leon et~al.(2021)Uribe-Leon, Vasquez, Giraldo, and
  Ricaurte}]{Uribe2021finding}
Uribe-Leon, C., Vasquez, J.~C., Giraldo, M.~A., and Ricaurte, G. (2021),
  \enquote{Finding Optimal Assortativity Configurations in Directed Networks,}
  \textit{Journal of Complex Networks}, 8, cnab004.

\bibitem[{van~der Hofstad(2017)}]{vdHofstad2017random}
van~der Hofstad, R. (2017), \textit{Random Graphs and Complex Networks},
  Cambridge, UK: Cambridge University Press.

\bibitem[{van~der Hoorn and Litvak(2015)}]{Van2015degree}
van~der Hoorn, P. and Litvak, N. (2015), \enquote{Degree-Degree Dependencies in
  Directed Networks with Heavy-Tailed Degrees,} \textit{Internet Mathematics},
  11, 155--179.

\bibitem[{Wan et~al.(2017)Wan, Wang, Davis, and Resnick}]{Wan2017fitting}
Wan, P., Wang, T., Davis, R.~A., and Resnick, S.~I. (2017), \enquote{Fitting
  the Linear Preferential Attachment Model,} \textit{Electronic Journal of
  Statistics}, 11, 3738--3780.

\bibitem[{Wan et~al.(2020)Wan, Wang, Davis, and Resnick}]{Wan2020extreme}
--- (2020), \enquote{Are Extreme Value Estimation Methods Useful For Network
  Data?} \textit{Extremes}, 23, 171--195.

\bibitem[{Wang and Resnick(2021{\natexlab{a}})}]{Wang2021common}
Wang, T. and Resnick, S.~I. (2021{\natexlab{a}}), \enquote{Common Growth
  Patterns for Regional Social Networks: {A} Point Process Approach,}
  \textit{Journal of Data Science}, https://doi.org/10.6339/21--JDS1021.

\bibitem[{Wang and Resnick(2021{\natexlab{b}})}]{Wang2021measuring}
--- (2021{\natexlab{b}}), \enquote{Measuring Reciprocity in a Directed
  Preferential Attachment Network,} \textit{Advances in Applied Probability},
  To appear.

\bibitem[{Yan et~al.(2021)Yan, Yuan, and Zhang}]{Rpkg:wdnet}
Yan, J., Yuan, Y., and Zhang, P. (2021), \textit{wdnet: {W}eighted {D}irected
  {N}etwork}, University of Connecticut, {R} package version 0.0-3,
  \url{https://gitlab.com/wdnetwork/wdnet}.

\bibitem[{Yuan et~al.(2021)Yuan, Yan, and Zhang}]{Yuan2021assortativity}
Yuan, Y., Yan, J., and Zhang, P. (2021), \enquote{Assortativity Measures for
  Weighted and Directed Networks,} \textit{Journal of Complex Networks}, 9,
  cnab017.

\end{thebibliography}

\appendix

\section{Interface with {\tt CVXR} Package}
\label{Append:express}

As mentioned, we use the utility functions from \code{CVXR} package 
to solve the optimization problem defined in 
Section~\ref{sec:rewiring}. The linear constraints of those 
functions are represented by vectors and matrices in the description 
file of \code{CVXR}. We hence write the constraints for our 
optimization problem in the form of matrices as well. Let $\bfk$ and 
$\bfl$ respectively be the collection of distinct out-degree and 
in-degree values in $G(V, E)$. Recall that we use $q_k^{(1)}$ to 
represent the probability that an edge emanates from a source node 
of out-degree $k$. In what follows, let the $\lvert\bfk\rvert$-long 
vector 
$\bfq^{(1)} := (q_{k}^{(1)})^\top$ denote the empirical out-degree 
distribution for source nodes, $\bftq^{(1)} := 
(\tq_{k}^{(1)})^\top$ denote the empirical out-degree distribution 
for target nodes,
where $\lvert\bfk\rvert$ is the cardinality of vector $\bfk$, 
and $\bfv^{\top}$ is the transpose of $\bfv$. In what follows, we 
define $\bfq^{(2)} := (q_{l}^{(2)})^\top$
and $\bftq^{(2)} := (\tq_{l}^{(2)})^\top$ in 
a similar manner. Consider two design matrices respectively given 
by $\bfR := \bfI_{\lvert\bfk\rvert \times \lvert\bfk\rvert} \otimes 
\bm{1}_{\lvert\bfl\rvert}^{\top}$ and $\bfS := 
\bm{1}_{\lvert\bfk\rvert}^{\top} 
\otimes \bfI_{\lvert\bfl\rvert \times \lvert\bfl\rvert}$, where 
$\bfI_{\lvert\bfk\rvert \times 
	\lvert\bfk\rvert}$ is a $\lvert\bfk\rvert \times 
	\lvert\bfk\rvert$ identity matrix, 
$\bm{1}_{\lvert\bfl\rvert}$ is an $\lvert\bfl\rvert$-long column 
vector consisting of 
all ones, and $\otimes$ represents \emph{Kronecker product}. Lastly, 
we use $\bfK$ to denote a $\lvert\bfk\rvert \times \lvert\bfl\rvert$ 
matrix, each column 
of which is $\bfk$. Analogously, $\bfL$ is defined as a 
$\lvert\bfl\rvert
\times \lvert\bfk\rvert$ matrix that is composed of $\bfl$'s. 

For the sake of implementation, we arrange all the quantities in 
$\bm{\eta}$ in the following matrix:
\begin{equation*}
	\bm{H} = 
	\begin{pmatrix}
		\eta_{k_1 l_1 k_1 l_1} & \eta_{k_1 l_1 k_1 l_2} & \cdots 
		& \eta_{k_1 l_1 k_{\lvert\bfk\rvert} l_{\lvert\bfl\rvert}} 
		\\ 
		\eta_{k_1 l_2 k_1 l_1} & \eta_{k_1 l_2 k_1 l_2} & \cdots 
		& \eta_{k_1 l_2 k_{\lvert\bfk\rvert} l_{\lvert\bfl\rvert}} 
		\\ 
		\vdots & \vdots & \ddots & \vdots \\ 
		\eta_{k_{\lvert\bfk\rvert} l_{\lvert\bfl\rvert} k_1 l_1} & 
		\eta_{k_{\lvert\bfk\rvert} 
			l_{\lvert\bfl\rvert} k_1 l_2} & \cdots & 
			\eta_{k_{\lvert\bfk\rvert} l_{\lvert\bfl\rvert} 
			k_{\lvert\bfk\rvert} l_{\lvert\bfl\rvert}}
	\end{pmatrix},
\end{equation*}
which is of dimension $(\lvert\bfk\rvert\lvert\bfl\rvert) \times 
(\lvert\bfk\rvert\lvert\bfl\rvert)$. We are 
now ready to rewrite the constraints for our convex optimization 
problems as follows:
\begin{align*}
	& \bm{H} \geq 0, \\
	& \bm{H} \bm{1}_{\lvert\bm{k}\rvert\lvert\bm{l}\rvert} = 
	\frac{{\rm \bf vec}\left((\bfK \odot \bm{\nu})^\top\right)}
	{\bm{1}_{\lvert\bfk\rvert}^{\top} (\bfK \odot \bm{\nu}) 
	\bm{1}_{\lvert\bfl\rvert}},
	\quad \bm{H}^\top \bm{1}_{\lvert\bm{k}\rvert\lvert\bm{l}\rvert} 
	= 
	\frac{{\rm \bf vec}\left(\bfL \odot \bm{\nu}^\top\right)}
	{\bm{1}_{\lvert\bfl\rvert}^{\top} (\bfL \odot \bm{\nu}^\top)
		\bm{1}_{\lvert\bfk\rvert}}, \\
	& r^*(1, 1) = \frac{\bfk^\top 
		\left(\bm{RHR}^\top - \bfq^{(1)}(\bftq^{(1)})^\top \right) 
		\bfk}
	{\sigma_q^{(1)}\sigma_{\tq}^{(1)}},\\
	& r^*(1, 2) = \frac{\bfk^\top 
		\left(\bm{RHS}^\top - \bfq^{(1)}(\bftq^{(2)})^\top \right) 
		\bfl}
	{\sigma_q^{(1)} \sigma_{\tq}^{(2)}},
	\\ 
	& r^*(2, 1) = \frac{\bfl^\top 
		\left(\bm{SHR}^\top - \bfq^{(2)}(\bftq^{(1)})^\top \right) 
		\bfk}
	{\sigma_q^{(2)} \sigma_{\tq}^{(1)}},\\
	& r^*(2, 2) = \frac{\bfl^\top 
		\left(\bm{SHS}^\top - \bfq^{(2)}(\bftq^{(2)})^\top\right) 
		\bfl}
	{\sigma_q^{(2)} \sigma_{\tq}^{(2)}},
\end{align*}
where $\odot$ represents element-wise product and ${\rm \bf 
	vec}(\cdot)$ is matrix vectorization operator.

\section{Probability Rules for Generating PA Networks}
\label{Append:PAgeneration}

The growth of PA networks is governed by a collection of parameters 
$\bftheta = (\alpha, \beta, \gamma, \deltain, \deltaout)$. Let $t 
\ge 0$ index the time, $G_{t + 1} (V_{t + 1}, E_{t + 1})$ is 
generated by adding an edge to $G_t (V_t, E_t)$ according to one of 
the following scenarios. Recall that $d_v^{(1)}$ and $d_v^{(2)}$ 
respectively represent the in- and out-degree of node $v$.
\begin{enumerate}
	\item With probability $\alpha$, a new directed edge $(v_1, 
	v_2)$ is added from a new node $v_1 \in V_{t + 1} \setminus V_t$ 
	to an existing node $v_2 \in V_t$, 
	where $v_2$ is chosen with probability 
	\begin{equation*}
		\Pr(\mbox{choose } v_2 \in V_t) = 
		\frac{\din_{v_2} + \deltain}
		{\sum_{v \in V_t} \left(\din_v + \deltain\right)};
	\end{equation*}
	\item With probability $\beta$, a new directed edge $(v_1, v_2)$ 
	is added between existing nodes from $v_1 \in V_{t + 1} = V_t$ 
	to $v_2 \in V_{t + 1} = V_t$, where $v_1$ and 
	$v_2$ are chosen independently with probability 
	\begin{equation*}
		\Pr(\mbox{choose } v_1, v_2 \in V_t) = 
		\left[\frac{\dout_{v_1} + \deltaout}
		{\sum_{v \in V_t} \left(\dout_v + \deltaout\right)}\right]
		\left[\frac{\din_{v_2} + \deltain}
		{\sum_{v \in V_t} \left(\din_v + \deltain\right)}\right];
	\end{equation*}
	\item With probability $\gamma$, a new directed edge $(v_1, v_2)$
	is added from an existing node $v_1 \in V_t$ to a new node 
	$v_2 \in V_{t + 1} \setminus V_t$, 
	where $v_1$ is chosen with probability 
	\begin{equation*}
		\Pr(\mbox{choose } v_1 \in V_t) = 
		\frac{\dout_{v_1} + \deltaout}
		{\sum_{v \in V_t} \left(\dout_v + \deltaout\right)}.
	\end{equation*}
\end{enumerate}

\section{Increases in Assortativity Coefficients under Different 
	Scenarios}
\label{Append:increase}

\begin{figure}[tbp]
	\centering
	\includegraphics[width = \textwidth]{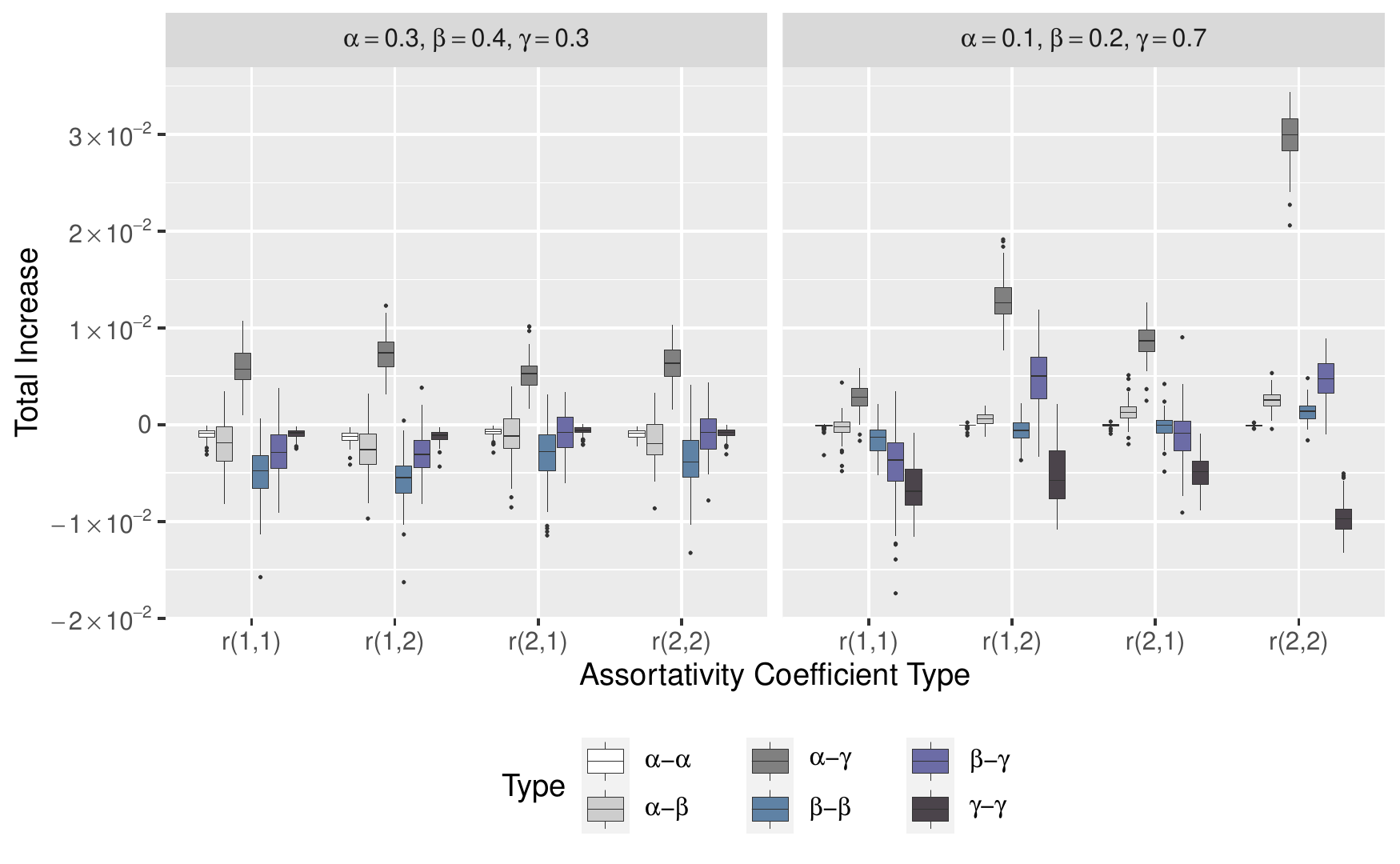}
	\caption{Side-by-side boxplots for the total increase in 
		assortativity coefficients under different rewiring scenarios
		from DPA networks with $\alpha = 0.3$, $\beta = 0.4$, 
		$\gamma = 
		0.3$ (left panel) and $\alpha = 0.1$, $\beta = 0.2$, $\gamma 
		= 
		0.7$ (right panel).} 
	\label{fig:PAincrease}
\end{figure}

Through simulation examples, we compare the increases in all four
assortativity coefficients for DPA networks contributed by different 
rewiring 
scenarios, i.e., $\alpha$-$\alpha$, $\alpha$-$\beta$, 
$\alpha$-$\gamma$, $\beta$-$\beta$, $\beta$-$\gamma$ and 
$\gamma$-$\gamma$ scenarios. In Figure~\ref{fig:PAincrease},
we present the results from: (1) $\alpha = 
0.3, \beta = 0.4, \gamma = 0.3$ under the $\alpha = \gamma$ setting 
and; (2) $\alpha = 0.1, \beta = 0.2, \gamma = 0.7$ under the setting 
of 
fixed $\beta$.

Though the chosen value of
$\beta$ in the first case (left panel) is larger, leading to a 
limited number of 
edges generated from the $\alpha$- and $\gamma$-scenarios, the 
contributions to the increases in all four types of assortativity 
coefficients by 
$\alpha$-$\gamma$ are still greater than the other combinations. In 
the second
case (right panel), we also see more contribution by 
$\alpha$-$\gamma$ when the value of $\beta$ decreases to 0.2. Note 
that the 
presented example is with the smallest value of $\alpha \gamma$ 
among all in the setting of fixed $\beta = 0.2$, so that a greater 
amount of increase in assortativity coefficients is expected when 
$\alpha 
\gamma$ gets large. Therefore, we conclude that the simulation 
results are in support of our elaborations in 
Section~\ref{sec:BAmodel}.

\end{document}